\documentclass[12pt]{article}
\usepackage{geometry}
\usepackage{fullpage}
\usepackage{microtype}
\usepackage[utf8]{inputenc}
\usepackage{setspace}

\usepackage{hyperref}
\usepackage[round]{natbib}
\usepackage{graphicx}
\usepackage[dvipsnames]{xcolor}
\usepackage{algorithm}
\usepackage{algpseudocode}
\usepackage{multirow}

\usepackage{amsmath, amssymb, amsthm}
\usepackage{amsfonts}

\newcommand{\E}{\mathbb{E}}
\newcommand{\I}{\mathbb{I}}
\newcommand{\var}{\operatorname{var}}
\newcommand{\cov}{\operatorname{cov}}
\newcommand{\cor}{\operatorname{cor}}

\newcommand{\dd}{\mathrm{d}}
\newcommand{\vecop}{\operatorname{vec}}

\newcommand{\diag}{\operatorname{diag}}

\newcommand{\tsp}{\mathsf{T}}

\newcommand{\rN}{\mathrm{N}}

\newcommand{\Beta}{\mathcal{B}}

\newcommand{\R}[1]{\mathbb{R}^{#1}}

\newcommand{\SPD}[1]{\mathbb{S}^{#1}_{++}}
\newcommand{\SPSD}[1]{\mathbb{S}^{#1}_{+}}
\newcommand{\DPD}[1]{\mathbb{D}^{#1}_{++}}

\DeclareMathOperator*{\argmin}{arg\,min}
\DeclareMathOperator*{\argmax}{arg\,max}

\theoremstyle{plain}
\newtheorem{theorem}{Theorem}[section]

\newtheorem{lemma}[theorem]{Lemma}

\theoremstyle{definition}

\newtheorem{example}{Example}
\theoremstyle{remark}

\title{Mixed-type multivariate response regression with covariance estimation}
\author{{\large Karl Oskar Ekvall$^*$ \hspace{3cm} Aaron J. Molstad$^\dagger$}\\
{\small \tt \hspace{-1cm} karl.oskar.ekvall@ki.se \hspace{3cm} amolstad@ufl.edu} \\
{\small Division of Biostatistics, Institute of Environmental Medicine,
Karolinska Institute$^*$}\\
{\small Applied Statistics Research Unit, Inst. of Stat. and Math. Methods in Econ., TU Wien$^*$}\\
{\small Department of Statistics and Genetics Institute, University of Florida$^\dagger$}
}
\date{{\small March 2022}}

\begin{document}

\maketitle

\begin{abstract}
  We propose a new method for multivariate response regression
  and covariance estimation when elements of the response vector are of mixed
  types, for example some continuous and some discrete. Our method is based on a
  model which assumes the observable mixed-type response vector is connected to a
  latent multivariate normal response linear regression through a link function.
  We explore the properties of this model and show its parameters are identifiable
  under reasonable conditions. We impose no parametric restrictions on the
  covariance of the latent normal other than positive definiteness, thereby
  avoiding assumptions about unobservable variables which can be difficult to
  verify in practice. To accommodate this generality, we propose a novel algorithm
  for approximate maximum likelihood estimation that works ``off-the-shelf" with
  many different combinations of response types, and which scales well in the
  dimension of the response vector. Our method typically gives better predictions
  and parameter estimates than fitting separate models for the different response
  types and allows for approximate likelihood ratio testing of relevant hypotheses
  such as independence of responses. The usefulness of the proposed method is
  illustrated in simulations; and one biomedical and one genomic data example.
\end{abstract}

\onehalfspacing

\section{Introduction}

In many regression applications, there are multiple response variables of
mixed types. For instance, when modeling complex biological processes like
fertility (see Section \ref{subsec:Fertility}), the outcome (ability to
conceive) is often best characterized through a collection of variables of
mixed types: both count (number of egg cells and number of embryos) and
continuous (square-root estradiol levels and log gonadotropin levels).
Similarly, one may be interested in the dependence between various types of
-omic data in a particular genomic region (see Section
\ref{subsec:SomaticMutations}), some binary (e.g., presence of somatic
mutations) and some continuous (e.g., gene expression). Joint modeling of
responses facilitates inference on their dependence, and it can lead to more
efficient estimation, better prediction, and allows for the testing of joint
hypotheses without the need for multiple testing corrections. Popular
regression models, however, typically assume all responses are of the same
type. A case in point is the multivariate normal linear regression model which
assumes a vector of responses $Y \in \R{r}$ and vector of predictors $x \in
\R{p}$ satisfy
\begin{equation}\label{eq:mlr}
  Y \sim \rN_r(\Beta^\tsp x, \Sigma),
\end{equation}
for some regression coefficient matrix $\Beta \in \R{p \times r}$ and
covariance matrix $\Sigma \in \SPD{r}$, the set of $r\times r$ symmetric and
positive definite matrices. Model \eqref{eq:mlr} is fundamental in
multivariate statistics and leads to relatively straightforward
likelihood-based inference; the parameters are identifiable, have intuitive
interpretations, and maximum likelihood estimates are computationally
tractable even when $r$ and $p$ are relatively large (but smaller than the
number of independent observations $n$). Here, our goal is the development of
a likelihood-based method for mixed-type response regression that retains some
of the useful properties of \eqref{eq:mlr}. We focus on settings where
dependence between responses cannot be parsimoniously parameterized and
interest is in inference on an unstructured $r\times r$ covariance matrix
characterizing this dependence. Rather than a method for specific combinations
of response distributions, of which there are many \citep{Olkin.Tate1961,
Poon.Lee1987, Catalano.Ryan1992, Cox.Wermuth1992, Fitzmaurice.Laird1995,
Catalano1997, Fitzmaurice.Laird1997, Gueorguieva.Agresti2001,
Gueorguieva.Sanacora2006, Yang.etal2007, Faes.etal2008}, we pursue a unified
method that can be adapted to many different combinations of response
distributions with relative ease.

To develop our method, we assume there is a known link function $g:\R{r}\to
\R{r}$ and latent vector $W \in \R{r}$ such that
\begin{equation}\label{eq:model}
    g\{\E(Y\mid W)\} = W ~~\text{and} ~~ W \sim \rN_r(\Beta^\tsp x, \Sigma).
\end{equation}
That is, a latent vector satisfies the multivariate linear regression model
and the observable responses are connected to that latent vector through
their conditional mean. We further suppose that, conditionally on $W$, the
elements of $Y$ satisfy generalized linear models with linear predictors given
by the elements of $W$ (see detailed specification in Section
\ref{sec:model}). This leads to a likelihood based on $n$ independent
observations
$\{(y_i, x_i) \in \R{r}\times \R{p}, i = 1, \dots, n\}$ which can be written
\begin{equation}\label{eq:lik}
  L_n(\Beta, \Sigma) = \prod_{i = 1}^n \int_{\R{r}} f(y_i\mid w_i)\, \phi (w_i; \Beta^\tsp x_i, \Sigma)\,\dd w_i,
\end{equation}
where $f(y_i\mid w_i)$ is the conditonal density for $Y_i = [Y_{i1}, \dots,
Y_{ir}]^\tsp$ given $W_{i} = [W_{i1}, \dots, W_{ir}]^\tsp$ and $\phi(\cdot;
\Beta^\tsp x_i, \Sigma)$ is the multivariate normal density with mean
$\Beta^\tsp x_i$ and covariance matrix $\Sigma$. Note \eqref{eq:mlr} is a
special case of \eqref{eq:model} where $Y = W$ and $g$ is the identity.
Conversely, \eqref{eq:lik} can be obtained as the likelihood for a specific
generalized linear latent and mixed model
(GLLAMM) \citep{Rabe-Hesketh.etal2004}, a class of models which also includes
many other models for mixed-type responses as special cases
\citep{Sammel.etal1997, Gueorguieva2001, deLeon.Carriegre2007,
Ekvall.Jones2020, Bai.etal2020, Kang.etal2021}. Example \ref{ex:one}
illustrates our setting.

\begin{example} \label{ex:one}
  Suppose $Y\in \R{20}$ is a vector of 10 count variables ($Y_1, \dots,
  Y_{10}$) and 10 continuous variables ($Y_{11}, \dots, Y_{20}$) whose joint
  distribution is of interest. If there are no predictors, a possible version
  of \eqref{eq:model} assumes $Y_{j} \mid W \sim \mathrm{Poi}(W_j)$ for $j =
  1, \dots, 10$, $Y_j \mid W \sim \rN(W_j, 1)$ for $j = 11, \dots, 20$, and $W
  \sim \rN_{20}(\Beta^\tsp, \Sigma)$, where $\Beta^\tsp \in \R{20}$ and
  $\Sigma \in \SPD{20}$.
\end{example}

There are several challenges with inference based on \eqref{eq:lik}. First, in
general the integrals cannot be computed analytically, and numerical
integration can be prohibitively time consuming even for $r \approx 10$. This
is commonly addressed by, for example, Laplace approximations, Monte Carlo
integration, or penalized quasi likelihood (PQL), which results from dropping
terms in a Laplace approximation that are assumed to be of lower order
\citep{Schall1991, Breslow.Clayton1993,Rabe-Hesketh.etal2002,
Breslow2004,Knudson.etal2021}. Here, in the interest of developing a method
operational for relatively large $r$, we will approximate the likelihood using
a modified version of PQL (Section \ref{sec:Computation}), which is known to
be fast \citep{Rabe-Hesketh.etal2002}. We will also discuss how the proposed
method can be extended to settings where the likelihood is approximated by
other means.

A second challenge is that, even if the integrals in \eqref{eq:lik}
can be computed efficiently, maximum likelihood estimation of $(\Beta, \Sigma)$
requires optimization over $\R{p\times r} \times \SPD{r}$:
\begin{equation}\label{eq:MLE}
  (\hat{\Beta}, \hat{\Sigma}) \in \argmax_{(\Beta, \Sigma) \in \R{p\times r}
  \times \SPD{r}} L_n(\Beta, \Sigma).
\end{equation}
When $r$ is small it may be possible to solve \eqref{eq:MLE} using
off-the-shelf optimizers (e.g., the {\tt ml} command in Stata or the {\tt
optim} function in R). For example, if $r = 2$, then \eqref{eq:MLE} can be
characterized as a constrained optimization problem over $(\Beta,
[\Sigma_{11}, \Sigma_{12}, \Sigma_{22}]^\tsp) \in \R{p\times r} \times \R{3}$,
the constraints being that $\Sigma_{11} > 0$, $\Sigma_{22} > 0$, and
$\Sigma_{12}^2 < \Sigma_{11} \Sigma_{22}$. When $r$ is even moderately large,
however, such constraints become untenable. In software packages for mixed and
latent variable models it is common to maximize the likelihood in $\Beta$ and
the Cholesky root $L$ such that $\Sigma =
LL^\tsp$ \citep{Rabe-Hesketh.etal2004a,Bates.etal2015}. This ensures
the estimate of $\Sigma$ is positive semi-definite, but it does
not ensure positive definiteness and it complicates more ambitious inference.
For example, implementing a likelihood-ratio test of whether some of the
responses are independent requires solving \eqref{eq:MLE} subject to the
constraint that some off-diagonal elements of $\Sigma$ are zero (Section
\ref{sec:testing}). Similarly, for some response-types it is necessary to
constrain diagonal elements of $\Sigma$ to ensure identifiability (see Section
\ref{sec:model}), and sometimes it is desirable to assume diagonal elements of
$\Sigma$ corresponding to responses of the same type (e.g., responses 1--10 or
11--20 in Example \ref{ex:one}) are the same. These types of restrictions are
non-trivial to accommodate when parameterizing in terms of $L$, but as we will
see in Section \ref{sec:Computation}, they are elegantly handled by the method
we propose. Briefly, our method iteratively updates a PQL-like approximation of
\eqref{eq:lik} and maximizes that approximation in $\Beta$ and $\Sigma$ using
block coordinate descent. The update for $\Beta$ is a least squares problem
which can be solved in closed form and the update for $\Sigma$ is solved using
an accelerated projected gradient descent algorithm. The algorithm scales well
in the dimension $r$ and it natively supports restrictions of the form $\Sigma
\in \mathbb{M}$ for sets $\mathbb{M} \subseteq \SPD{r}$ onto which a
projection can be computed.

In some settings it may be appropriate to, as is common in mixed models,
restrict $\Sigma = Z DZ^\tsp$ for some design matrix $Z \in \R{r\times d}$ and
covariance matrix $D \in \SPD{d}$ of $d < r$ random effects \citep[see e.g.][]{Jiang2007}. Such structures may be suggested by subject-specific
knowledge or the sampling design, or they may be motivated by necessity when
$n$ is small relative to $r$. Similarly, many methods based on the marginal
moments, such as for example generalized estimating equations
\citep{Liang.Zeger1986}, specify a covariance structure. In many applications,
however, it is unclear whether there is any particular dependence structure.
Indeed, it may be of primary interest to discover a structure using data,
rather than imposing one {\it a priori}; our method enables such discoveries.
Some methods based on the marginal moments also allow the specification of an
unstructured correlation matrix, but for $Y$ rather than
$W$ \citep{Rochon1996,Bonat.Jorgensen2016}. In general, however, such methods
are inconsistent with \eqref{eq:model} since the mean vector and correlation
matrix of $Y$ cannot be independently parameterized. Accordingly, it is often
unclear whether the estimates are in the parameter set of any particular model
for mixed-type responses. Nevertheless, methods based on the marginal moments
can be useful in practice and they are a reasonable comparison to our
method for some purposes.

Other advantages of the parameterization we consider are that (i) the off-diagonal
elements of $\Sigma$ affect the joint distribution of responses but not their
marginal (univariate) distributions and (ii) responses are independent if the
corresponding off-diagonal elements of $\Sigma$ are zero. Thus, we can test the
null hypothesis that some responses are independent by testing whether the
corresponding off-diagonal elements of $\Sigma$ are zero, without that null
hypothesis implying restrictions on the marginal distributions. This is
generally not possible in more parsimonious parameterizations of the form
$\Sigma = ZDZ^\tsp$ since, in general, elements of $D$ affect both marginal
and joint distributions.

\section{Model}\label{sec:model}
\subsection{Specification}
  We now specify our model in detail and generalize somewhat compared to the
  Introduction. Assume the elements of $Y = (Y_1, \dots, Y_r)$ are
  conditionally independent given $W$ with conditional densities of the form
  \begin{equation}\label{eq:uni_cond_dist}
    f(y_j\mid w) = \exp\left\{\frac{y_j w_j - c_j(w_j)}{\psi_j}\right\},
  \end{equation}
  where $\psi = [\psi_1, \dots, \psi_r]^\tsp$ is a vector of strictly positive
  dispersion parameters and $c_j$ is a cumulant function for the distribution
  of $Y_j \mid W$. For example, $c_j'(W_j) = \E(Y_j\mid W)$ and $\psi_j
  c_j''(W_j) = \var(Y_j \mid W)$, with primes denoting derivatives. From these
  properties it follows that the link function $g$ defined by $g\{\E(Y\mid W\}
  = W$ satisfies $g(W) = [g_1(W_1), \dots, g_r(W_r)]^\tsp$ with $g_j$
  real-valued and strictly increasing for $j = 1, \dots, r$. That is, the $j$th latent
  variable has a direct effect on the $j$th response but no other responses.
  Equation \eqref{eq:uni_cond_dist} specifies a (conditional) generalized
  linear model (GLM) \citep{McCullagh.Nelder1989}, which when $\psi_j = 1$
  specializes to a one-parameter exponential family distribution, as in
  Example \ref{ex:one}. Here, we do not assume $\psi_j = 1$ for every $j$ but
  we do assume the $\psi_j$ are known.

  Because it makes our development no more difficult, in what follows
  we consider a slightly more general version of \eqref{eq:model} where
  \begin{equation} \label{eq:latent_reg}
    W \sim \rN_r(X\beta, \Sigma)
  \end{equation}
  for a non-stochastic design matrix $X \in \R{r\times q}$ and $\beta \in
  \R{q}$. The classical multivariate response regression setting in
  \eqref{eq:model} which motivates our study is then a special case with $X =
  I_r \otimes x^\tsp$, $\beta = \vecop(\Beta)$, and $q = rp$, where $\otimes$
  is the Kronecker product and $\vecop(\cdot)$ the vectorization operator
  stacking the columns of its matrix argument. Unlike \eqref{eq:model} where
  all responses have the same predictors, \eqref{eq:latent_reg} allows
  distinct predictors for each response, as in seemingly unrelated
  regressions \citep{Zellner1962}.

  With $\{(y_i, X_i) \in \R{r} \times \R{r \times p}, i = 1, \dots, n\}$,
  denoting independent realizations, the likelihood is
  \begin{equation} \label{eq:lik_detailed}
    L_n(\beta, \Sigma) = \vert \Sigma\vert^{-n/2} \prod_{i = 1}^n \int_{\R{r}}
    \exp\left\{\left(\sum_{j = 1}^r \frac{y_{ij} w_{ij} - c_j(w_{ij})}{\psi_j}\right) -
    \frac{1}{2}(w_i - X_i \beta)^\tsp \Sigma^{-1}(y_i - X_i\beta)\right\} \dd
    w_i.
  \end{equation}
  To see the connection to GLLAMMs and other mixed models which are often
  written for all observations simultaneously, let $\mathcal{Y} = [Y_1^\tsp,
  \dots, Y_n^\tsp]^\tsp \in \R{rn}$, $\mathcal{X} = [X_1^\tsp, \dots,
  X_n^\tsp]^\tsp \in \R{rn \times q}$, and $\mathcal{E} \sim \rN_{nr}(0, I_n
  \otimes \Sigma)$. Then \eqref{eq:lik_detailed} is the likelihood for a
  model which assumes the elements of $\mathcal{Y}$ follow conditionally
  independent GLMs given $\mathcal{E}$, with canonical link functions and
  linear predictors given by the elements of
  \[
    \mathcal{W} = \mathcal{X}\beta + \mathcal{E};
  \]
  that is, each of the $rn$ responses has a linear predictor with its own
  random intercept, and the covariance matrix for those random intercepts is
  $I_n \otimes \Sigma$.

  \subsection{Parameter interpretation and identifiability}

  It is often difficult to interpret parameters in latent variable models and,
  similarly, it is often unclear which parameters are identifiable -- we
  address some such concerns in this section. The parameters are
  straightforward to interpret in the latent regression, but interpreting them
  in the marginal distribution of $Y$ requires more work. To that end, note
  that the mean vector and covariance matrix of $Y$ are, respectively, by
  iterated expectations,
  \begin{equation} \label{eq:mean:cov}
    \E(Y) = \E\{g^{-1}(W)\}~~\text{and}~~\cov(Y) = \cov\{g^{-1}(W)\} +
    \E\{\cov(Y\mid W)\}.
  \end{equation}
  We make a number of observations based on \eqref{eq:mean:cov}: first, because
  $\cov(Y\mid W)$ is assumed diagonal, the covariance between responses
  is determined by $\cov\{g^{-1}(W)\}$. Second, since $\E(Y_j)$ and $\E(Y_j^2)$
  are determined by the univariate distribution of $Y_j$, off-diagonal
  elements of $\Sigma$ do not affect means and variances of the responses.
  Third, since $g$ and $\cov(Y\mid W)$ are non-linear and non-constant in
  general, $\E(Y)$ and $\E\{\cov(Y\mid W)\}$ in general depend on both $\beta$
  and diagonal elements of $\Sigma$. Fourth, since $\var(Y_j)$ is increasing
  in $\psi_j$ and $\cov\{g^{-1}(W)\}$ does not depend on $\psi$, $\cor(Y_j,
  Y_k)$ is decreasing in $\psi_j$ and $\psi_k$. This is intuitive as
  responses are conditionally uncorrelated and hence, loosely speaking, a
  large element of $\psi$ indicates substantial variation in the corresponding
  response is independent of the variation in the other responses. In some
  settings, more precise statements are possible by analyzing closed form
  expressions for the moments in \eqref{eq:mean:cov}, as the next example
  illustrates.

    \begin{example} \label{ex:norm:poi}
    (Normal and Poisson responses) Suppose there are $r = 4$ responses such
    that $\E(Y_j \mid W) = W_j$ and $\var(Y_j \mid W) = \psi_j$ for $j = 1,
    2$, and $\E(Y_j \mid W) = \exp(W_j)$ and $\var(Y_j \mid W) = \psi_j
    \exp(W_j)$ for $j = 3, 4$. These moments are consistent with assuming $Y_j
    \mid W \sim \rN(W_j, \psi_j)$ for $j = 1, 2$, and, if $\psi_3 = \psi_4 =
    1$, $Y_j \mid W \sim \mathrm{Poi}\{\exp(W_j)\}$, $j = 3, 4$. When not
    assuming $\psi_3 = \psi_4 = 1$, we say these moments are consistent with
    normal and (conditional) quasi-Poisson distributions. We examine the
    effects of these assumptions on the marginal moments of $Y$. Some algebra (Supplementary Materials)
    gives the following moments: $\E(Y_1) = X_1^\tsp
    \beta$, $\E(Y_3) = \exp(X_3^\tsp \beta + \Sigma_{33} / 2)$, $\var(Y_1) =
    \psi_1 + \Sigma_{11}$, $\var(Y_3) = \exp(2X_3^\tsp \beta +
    \Sigma_{33})\{\exp(\Sigma_{3 3}) - 1 + \psi_3 \exp(-X_3^\tsp\beta -
    \Sigma_{33}/2)\}$, $\cov(Y_1, Y_2) = \Sigma_{21}$, $\cov(Y_1, Y_3) =
    \Sigma_{31}\exp(X_3^\tsp \beta + \Sigma_{33}/2)$, and $\cov(Y_3, Y_4) =
    \exp(X_3^\tsp\beta + X_4^\tsp\beta + \Sigma_{33}/2 + \Sigma_{44} /
    2)\{\exp(\Sigma_{43}) - 1\}$; the remaining entries of $\cov(Y)$ are the
    same as those given up to obvious changes in subscripts. Clearly, both
    $\E(Y)$ and $\cov(Y)$ depend on $\beta$ and $\Sigma$, but regardless of
    type, the variance of $Y_j$ is increasing in $\Sigma_{jj}$, the mean is
    increasing in $X_j^\tsp \beta$, and the covariance between $Y_j$ and $Y_k$
    is increasing in $\Sigma_{jk}$. We will later use these observations to
    prove a result which implies $\beta$ and $\Sigma$ are identifiable in this
    example.

    Consider the linear dependence between responses with conditional normal
    and quasi-Poisson moments, $Y_1$ and $Y_3$, say. The sign of their
    correlation is the sign of $\Sigma_{13}$ and the squared correlation
    satisfies, by Cauchy--Schwarz's inequality, $ \cor(Y_1, Y_3)^2 \leq
    \Sigma_{1 1} \Sigma_{3 3} / [(\psi_1 + \Sigma_{1 1})\{\exp(\Sigma_{3 3}) -
    1 + \psi_3 / \E(Y_3)\}] \leq \Sigma_{3 3} / \{\exp(\Sigma_{3 3}) - 1\}$.
    Thus, strong linear dependence between $Y_1$ and $Y_3$ requires a small
    $\Sigma_{3 3}$.

    To gain intuition for how two responses with quasi-Poisson moments
    behave, suppose for simplicity that $\Sigma_{33} = \Sigma_{44}$, $\psi_3 =
    \psi_4$, and $X_3^\tsp \beta = X_4^\tsp \beta$. Then $\cor(Y_3, Y_4) =
    \{\exp(\Sigma_{43}) - 1\}\{\exp(\Sigma_{33}) - 1 + \psi_3 / \E(Y_3)\}$.
    For small $\psi_3$, this correlation is approximately $(\exp(\Sigma_{4 3})
    - 1) / (\exp(\Sigma_{33}) - 1)$, which for $\vert \Sigma_{43}\vert \leq
    \Sigma_{33}$ is upper bounded by $1$ and lower bounded by
    $\{\exp(-\Sigma_{33}) - 1\} / \{\exp(\Sigma_{33}) - 1\}$. The latter
    expression tends to $-1$ if $\Sigma_{3 3} \to 0$ and $0$ if $\Sigma_{3 3}
    \to \infty$. Thus, strong negative correlation between $Y_3$ and $Y_4$
    requires a small $\Sigma_{33}$.
  \end{example}

  Example \ref{ex:norm:poi} is convenient because the moments have
  closed form expressions. In more complicated settings, the following result
  can be useful. It implies the mean of $Y_j$ and covariance of $Y_j$ and
  $Y_k$ are strictly increasing in, respectively, the mean of $W_j$ and
  covariance between $W_j$ and $W_k$.

  \begin{lemma} \label{lem:cov}
    Let $\phi_{\mu, \Sigma}$ be a bivariate normal density with marginal
    densities $\phi_{\mu_1, \sigma_1^2}$ and $\phi_{\mu_2, \sigma_2^2}$ and
    covariance $\sigma = \Sigma_{12} = \Sigma_{21}$; then for any increasing,
    non-constant $g, h : \R{} \to \R{}$, the functions defined by $\mu_1
    \mapsto \int g(t) \phi_{\mu_1, \sigma_1^2}(t)\, \dd t$ and $\sigma \mapsto
    \int g(t_1)h(t_2)\phi_{\mu, \Sigma}(t)\, \dd t$ are, assuming the
    (Lebesgue) integrals exist, strictly increasing on $\R{}$ and
    $(-\sqrt{\Sigma_{11}\Sigma_{22}}, \sqrt{\Sigma_{11}\Sigma_{22}})$,
    respectively.
  \end{lemma}

  The proof is in the Supplementary Materials. We illustrate the usefulness of
  this result in another example.

\begin{example}[Normal and Bernoulli responses] \label{ex:bern}
Suppose $r = 2$ with $Y_1 \mid W_1 \sim \rN(W_1, \psi_1)$ and $Y_2$ Bernoulli
distributed with $ \E(Y_2 \mid W_2) = \mathrm{logit}^{-1}(W_2)$ $ =1 / \{1 +
\exp(- W_2)\}$. Suppose also for simplicity $W\sim \rN_2(\beta, \Sigma)$, $\beta \in
\R{2}$. The marginal distribution of $Y_2$ is Bernoulli with $ \E(Y_2) = \int
[\phi(t) / \{1 + \exp(- \beta_2 - \sqrt{\Sigma}_{22} t)\}]\,\dd t,$ where
$\phi(\cdot)$ is the standard normal density. One can show that, if
$\Sigma_{22}$ is fixed, $\E(Y_2) \to 0$ if $\beta_2 \to -\infty$ and $\E(Y_2)
\to 1$ if $\beta_2 \to \infty$. That is, any success probability is attainable
by varying $\beta$ and, hence, some restrictions are needed for
identifiability. One possibility, which has been used in similar settings, is
to fix $\Sigma_{22}$ to some value, say one \citep{Dunson2000,
Bai.etal2020}. While fixing $\Sigma_{22} = 1$ does not impose restrictions
on the distribution of $Y_2$ as long as $\beta_2$ can vary freely, it may
impose restrictions on the joint distribution of $Y = [Y_1, Y_2]^\tsp$,
properties of which we consider next.

Equation \eqref{eq:mean:cov} implies $\cov(Y_1, Y_2) =  \int [\{t_1
\phi_{\beta, \Sigma}(t)\} / \{1 + \exp(-t_2)\}]\,\dd t - \beta_1 \E(Y_2)$. The
integral does not admit a closed form expression, but Lemma \ref{lem:cov} says
the covariance is strictly increasing in $\Sigma_{12}$, which can be used to
show the parameters are identifiable in this example if $\Sigma_{22}$ is known
(Theorem \ref{thm:id}). To understand which values $\cov(Y_1, Y_2)$ can take,
consider the limiting case as $\Sigma_{12} \to
\sqrt{\Sigma}_{11}\sqrt{\Sigma}_{22}$ and assume for simplicity $\beta_1 =
\beta_2 = 0$. In the limit, the covariance matrix is singular and the
distribution of $W$ the same as that obtained by letting $W_2 =
(\sqrt{\Sigma_{2 2}} / \sqrt{\Sigma_{11}}) W_1$. Then one can show $ \cor(Y_1,
Y_2) =  2 \int [\{\sqrt{\Sigma_{11}} t\phi(t)\} / \{1 +
\exp(-\sqrt{\Sigma_{22}}t)\}]\, \dd t / \sqrt{\psi_1 + \Sigma_{11} }$. By
using the dominated convergence theorem as $\psi_1 \to 0$ and $\Sigma_{22} \to
\infty$, this can be shown to tend to and be upper bounded by $\sqrt{2 / \pi}
\approx 0.8$. This correlation corresponds to a limiting case and is an upper
bound on the attainable correlation between Bernoulli and normal responses.
\end{example}

We conclude with a result on identifiability. The result is stated for some
common choices of link functions and distributions of $Y\mid W$, but the proof
can be adapted to other settings. In the Supplementary Materials, we state a
result (Lemma B.4) which outlines conditions for identifiability more generally.
Essentially, when $Y_j\mid W$ satisfies a GLM, it suffices that, for every $j$,
the variance of $Y_j$ is not a function of the mean of $Y_j$. If it is, as is
the case when $Y_j$ is Bernoulli distributed, restrictions on diagonal elements
of $\Sigma$ are needed for identifiability. The proof is in the Supplementary
Materials.

\begin{theorem} \label{thm:id}
Suppose $\{(y_i, X_i) \in \R{r} \times \R{r\times q}; i = 1, \dots, n\}$ is an
independent sample from our model and that (i) $g_j$, $j = 1, \dots, r$, is
either the identity, natural logarithm, or logit (log-odds) function; (ii)
$\Sigma_{jj}$ is fixed and known for every $j$ corresponding to logit $g_j$;
and (ii) $\mathcal{X}^\tsp \mathcal{X}$ is invertible, then distinct $(\beta,
\Sigma)$ correspond to distinct distributions for $\mathcal{Y}$.
\end{theorem}

From a computational perspective, non-identifiability can lead to
likelihoods with infinitely many maximizers or ridges along which the likelihood
is constant. In light of this result, we constrain the diagonals of $\Sigma$
corresponding to binary response variables: we found this leads to faster
computation and improved estimation accuracy.

\section{Estimation}\label{sec:Computation}

\subsection{Overview}

We propose an algorithm based on linearization of the conditional mean function
$w\mapsto \nabla c(w) = g^{-1}(w)$, where $c(w) = \sum_{j = 1}^r c_j(w_j)$; this is essentially equivalent to
linearization of the link function, which has been considered in other latent
variable models \citep{Schall1991}. More specifically, consider the elementwise
first order Taylor approximation of $g^{-1}(\cdot) = \nabla c(\cdot)$ around an
arbitrary $w\in \R{r}$: $\E(Y\mid W) = \nabla c(W) \approx \nabla c(w) +
\nabla^2 c(w)(W - w)$. Applying expectations and covariances on both sides
yields $\E(Y) \approx \nabla c(w) + \nabla^2 c(w)(X\beta - w) =:  m(w, \beta)$
and $\cov\{\E(Y\mid W)\} \approx \nabla^2 c(w) \Sigma \nabla^2 c(w)$.
Approximating $\E\{\cov(Y\mid W)\} = \E\{\diag(\psi) \nabla^2 c(W)\} \approx
\diag(\psi) \nabla^2 c(w)$ leads to $\cov(Y) \approx \diag(\psi)\nabla^2 c(w) +
\nabla^2 c(w) \Sigma \nabla^2 c(w) =: C(w, \Sigma)$. Intuitively, we expect
$m(w, \beta)$ and $C(w, \Sigma)$ to be good approximations if $W$ takes values
near $w$ with high probability. Now, consider a working model which says $Y_1,
\dots, Y_n$ are independent with
\begin{equation} \label{eq:work:model}
Y_i \sim \rN_r \{m(w_i, \beta), C(w_i, \Sigma)\},
\end{equation}
for observation-specific approximation points $w_i \in \R{r}$, $i = 1, \dots,
n$. The corresponding negative log-likelihood is, up to scaling and additive
constants
\[
h_n(\beta, \Sigma\mid w_1, \dots, w_n) = \sum_{i = 1}^n\log {\rm
det}\{ C(w_i, \Sigma)\} +  \sum_{i = 1}^n\{y_i - m(w_i, \beta)\}^\tsp C(w_i,
\Sigma)^{-1}\{y_i - m(w_i, \beta)\}.
\]
If all responses are normal, then the working model is exact and minimizers of
$h_n$ are maximum likelihood estimates (MLEs). More generally, minimizers of
$h_n$ are approximate MLEs whose quality depend on the accuracy of the working
model \eqref{eq:work:model}. For further insight it is helpful to note that if
$w_i$ is set to the maximizer of the $i$th integrand in \eqref{eq:lik_detailed},
then $h_n$ is the same type of approximation of $L_n$ as that used in PQL,
specialized to our setting. The correspondence between linearization of the link
function and PQL is detailed by \citet{Breslow2004}. The correspondence
implies that, in addition to the moment-based motivation given here, $h_n$ can
be motivated as a Laplace approximation of $L_n$, with terms assumed to be of
lower stochastic order ignored \citep[see][for details]{Breslow.Clayton1993}. Roughly speaking, the approximation
will be better the closer the distribution of $Y_i$ is to a normal. Because the
latent variables are normal, we expect the distribution of $Y_i$ to be close to
normal if the distribution of $Y_i \mid W_i$ is. Now, to estimate variance
parameters, conventional PQL makes further approximations which lead to a set of
estimating equations. We proceed differently and avoid these
approximations, both because they lack formal motivation \citep[Section 2.5]{Breslow.Clayton1993} and because they do not in general lead to a
simpler optimization problem for an unstructured and positive definite $\Sigma$.
Thus, we will work directly with the approximate log-likelihood $h_n$ and
propose an algorithm that is substantially different from ones commonly used for
mixed models.

With $h_n$ as the starting point, a natural algorithm for estimating $\beta$ and
$\Sigma$ would iterate between updating $(\beta, \Sigma)$ by minimizing $h_n$
with the $w_i$ held fixed and then updating the $w_i$ to get a more accurate
working model. Motivated by the connection to Laplace approximations, we
update the $w_i$ by setting them to equal to the maximizers of the integrand in
\eqref{eq:lik_detailed} with $\beta$ and $\Sigma$ fixed at their current
iterates. To summarize, we propose a blockwise iterative algorithm whose
$(k+1)$th iterates are obtained using the updating equations
\begin{align}
(\beta^{(k+1)},\Sigma^{(k+1)}) &= \argmin_{\beta, \Sigma} h_n(\beta, \Sigma
\mid w_1^{(k)}, \dots, w_n^{(k)}), \label{eq:Sigma_beta_update}\\
(w_1^{(k+1)}, \dots, w_n^{(k+1)}) &= \argmax_{w_1, \dots, w_n} \sum_{i=1}^n
\log f_{\beta^{(k + 1)}, \Sigma^{(k + 1)}}(y_i, w_i).\label{eq:W_update}
\end{align}
This algorithm can be run for a pre-specified number of iterations or until
convergence of the $\beta$ and $\Sigma$ iterates, for example. While the
complete algorithm is not designed to minimize a particular objective function,
the individual updates, which we discuss in more detail shortly, minimize
objective functions that can be tracked to determine convergence within each
update. In our experience, the values of $\Sigma$ and $\beta$ tend to converge
after (at most) tens of iterations of \eqref{eq:Sigma_beta_update} and
\eqref{eq:W_update}. The final iterates of $\Sigma$ and $\beta$ are approximate
MLEs and $h_n$ evaluated at the final iterates of $w_1,\dots, w_n$ is an
approximate log-likelihood which we will use for approximate likelihood-based
inference, including the construction of likelihood-ratio tests.

A formal statement of the proposed algorithm is in Algorithm 1.

\begin{algorithm}[t]
\noindent \textbf{Algorithm 1}: Blockwise iterative algorithm for estimating $(\beta, \Sigma)$
\begin{enumerate}
\item Given $\epsilon_\beta > 0$, $\epsilon_\Sigma > 0$, initialize $\Sigma^{(1)} \in \mathbb{M}$, and $\beta^{(1)} \in
\mathbb{R}^{q}.$ Set $k = 1$.
\item $w_i^{(k+1)} = \argmax_{w \in \mathbb{R}^r} \left\{\log
f_{\beta^{(k)}, \underline{\Sigma}^{(k)}}(y_i, w) - \tau \Vert y_i -
X_i\beta^{(k)}\Vert^2\right\}$ for $i = 1, \dots, n$.
\item Set $\tilde{\Sigma}^{(1)} =
\Sigma^{(k)}$. For $l = 1, 2, \dots$ until convergence:
\begin{enumerate}
  \item $\tilde{\beta}^{(l + 1)} = \argmin_\beta h_n(\beta,
  \tilde{\Sigma}^{(l)} \mid w_1^{(k+1)}, \dots, w_n^{(k+1)})$
  \item Set $\bar{\Sigma}^{(0)} = \bar{\Sigma}^{(1)} = \tilde{\Sigma}^{(t)}$.
  For $t = 1, 2, \dots,$ until convergence:
  \begin{enumerate}
      \item[] $\bar{\Sigma}^{(t+1)} = \mathcal{P}_{\mathbb{M}}\left[
      \bar{\Sigma}^{(t)} - \alpha \nabla_\Sigma h_n(\tilde{\beta}^{(l+1)},
      \bar{\Sigma}^{(t)}, w_1^{(k+1)}, \dots, w_n^{(k+1)}) + \gamma \{
      \bar{\Sigma}^{(t)} - \bar{\Sigma}^{(t-1)}\} \right],$
  \end{enumerate}
  \item $\tilde{\Sigma}^{(l+1)} = \bar{\Sigma}^{(t^*)}$ where
  $\bar{\Sigma}^{(t^*)}$ is the final iterate from 3(b).
  \end{enumerate}
\item $(\beta^{(k+1)}, \Sigma^{(k+1)}) = (\tilde{\beta}^{(t^*)},
\tilde{\Sigma}^{(t^*)})$ where $(\tilde{\beta}^{(l^*)}, \tilde{\Sigma}^{(l^*)})$
are the final iterates from 3.
\item If $\|\beta^{(k+1)} - \beta^{(k)}\|_F^2 \leq \epsilon_\beta$ and
$\|\Sigma^{(k+1)} - \Sigma^{(k)}\|_F^2 \leq \epsilon_\Sigma$, terminate.
Otherwise, set $k \leftarrow k + 1$ and return to 2.
\end{enumerate}
\end{algorithm}


\subsection{Updating $\beta$ and $\Sigma$}
To solve \eqref{eq:Sigma_beta_update}, we use a
blockwise coordinate descent algorithm. Treating $w_1, \dots,
w_n$ as fixed throughout and ignoring the iterate
superscript, this algorithm iterates between updating $\beta$ and
$\Sigma$. Specifically, the $(l+1)$th iterates of the
algorithm for solving \eqref{eq:Sigma_beta_update} can be expressed
\begin{align}
\beta^{(l+1)} &= \argmin_{\beta} h_n(\beta, \Sigma^{(l)} \mid w_{1}, \dots,
w_n),\label{eq:beta_update}\\
\Sigma^{(l+1)} &= \argmin_{\Sigma} h_n(\beta^{(l+1)}, \Sigma \mid w_{1},
\dots, w_n).\label{eq:Sigma_update}
\end{align}
Update \eqref{eq:beta_update} can be shown to be a weighted residual
sum-of-squares with solution
\[
\beta^{(l+1)} = \left\{\sum_{i = 1}^n \tilde{X}_i^\tsp C(w_i,
\Sigma^{(l)})^{-1}\tilde{X}_i \right\}^{-1}\sum_{i = 1}^n \tilde{X}_i^\tsp
C(w_i, \Sigma^{(l)})^{-1}\tilde{y}_i,
\]
where $\tilde{X}_i = \nabla^2 c(w_i) X_i$ and $\tilde{y}_i = y_i - \nabla c(w_i)
+ \nabla^2 c(w_i)w_i$. Minimizing $h_n$ with respect to $\Sigma$ is non-trivial
owing to non-convexity and the constraint that $\Sigma$ is positive
semi-definite. One possibility is to parameterize $\Sigma$ in a way that lends
itself to unconstrained optimization \citep[e.g.][]{Pinheiro.Bates1996} and use a generic solver. However, such
parameterizations are inconvenient since we, as discussed in Section
\ref{sec:model}, sometimes restrict diagonal elements of $\Sigma$ to be equal to
a prespecified constant for identifiability. Similarly, testing correlation of
responses requires constraining some off-diagonal elements to equal zero. Thus,
we need an algorithm that allows restrictions on the elements of $\Sigma$
and ensures estimates are positive semi-definite, and can be constrained
to be positive definite if desirable.

By picking an appropriate (convex) $\mathbb{M} \subseteq \R{r\times r}$,
\eqref{eq:Sigma_update} can be characterized as an optimization problem over
$\R{r\times r}$ with the constraint that $\Sigma \in \mathbb{M}$. To handle both
the non-convexity and general constraints, we propose to solve this problem
using a variation of the inertial proximal algorithm proposed by \citet{Ochs.etal2014}. This is an accelerated projected gradient descent
algorithm that can be used to minimize an objective function which is the sum of
a non-convex smooth function and convex non-smooth function. In our case, $h_n$
(as a function of $\Sigma$) is the non-convex smooth function and the convex
non-smooth function is the function which equals $\infty$ if $\Sigma \not\in
\mathbb{M}$ and zero otherwise. This algorithm, like many popular accelerated
first order algorithms, e.g., FISTA \citep{Beck.Teboulle2009}, uses ``inertia''
in the sense that the search point is informed by the direction of
change from the previous iteration, which can lead to faster convergence.

To summarize briefly, our algorithm for \eqref{eq:Sigma_update} has $(t + 1)$th
iterate $\Sigma^{(t+1)} = \mathcal{P}_{\mathbb{M}}[ \Sigma^{(t)} - \alpha
\nabla_\Sigma h_n(\beta, \Sigma^{(t)} \mid w_1, \dots, w_n) + \gamma \{
\Sigma^{(t)} - \Sigma^{(t-1)}\}]$, where $\gamma \in (0,1)$, $\alpha$ is
determined using backtracking line search \citep[see][Algorithm 4]{Ochs.etal2014}, and $\mathcal{P}_{\mathbb{M}}$ is the projection onto
$\mathbb{M}$. We assume the projection is defined; it suffices, for example,
that $\mathbb{M}$ is non-empty, closed, and convex \citep[Corollary 5.1.19]{Megginson1998}. In our software, $\mathbb{M}$ is the
intersection of a set of matrices with constrained elements and the set of
symmetric matrices with eigenvalues bounded below by $\epsilon \geq 0$, where
$\epsilon = 0$ ensures positive semi-definiteness and $\epsilon > 0$ positive
definiteness. To compute projections onto $\mathbb{M}$ of this form, we
implement Dykstra's alternating projection algorithm \citep{Boyle.Dykstra1986}.
This algorithm iterates between projections onto each of the two sets whose
intersection defines $\mathbb{M}$. Both projections can be computed in closed
form, so this algorithm tends to be very efficient. The gradient of $h_n$ with
respect to $\Sigma$ needed for implementing the algorithm can be found in the
Supplementary Materials. The algorithm is terminated when the objective
function values converge.

\subsection{Updating the approximation points}
We use a trust region algorithm for updating $w_i, i = 1, \dots, n$ \citep[see e.g.][Chapter 4]{Nocedal.Wright2006}. Essentially, the trust region
algorithm approximates the objective function locally by a quadratic and
requires the computation of gradients and Hessians. The gradient is given in the
Supplementary Materials and the Hessian is, assuming $\Sigma^{-1}$ is positive
definite, for $i = 1, \dots, n$, $\nabla^2_{w_i} \log f_\theta(y_i, w_i) = -
\nabla^2 c(w_i) - \Sigma^{-1}$. Since $\nabla^2 c(w_i)$ and $\Sigma^{-1}$ are
positive definite and the latter does not depend on $w_i$, the objective
function is strongly concave and therefore has a unique maximizer and stationary
point. In practice, however, $\Sigma$ can be singular or near-singular and the
Hessian $-\Sigma^{-1} - \nabla^2c(w)$ can have a large condition number. To
improve stability, we regularize by (i) adding an $L_2$-penalty on $w_i -
X_i\beta$ and (ii) replacing $\Sigma$ by $ \underline{\Sigma} = \Sigma +
\kappa I_r$ for some small $\kappa > 0$. Then the optimization problem for
updating $w_i$ is
\[
\argmin_{w \in \R{r}}\left\{-y_i^\tsp w + c(w) + \frac{1}{2}(w_i -
X_i^\tsp \beta)^\tsp \underline{\Sigma}^{-1}(w_i - X_i\beta)  + \tau \Vert w_i - X_i^\tsp
\beta\Vert^2 \right\},
\]
where $\tau \geq 0$. The intuition for shrinking $w_i$ to $X_i \beta$ is that
the latter is the mean of $W_i$ when $\beta$ is the true parameter. The penalty
and regularization of $\Sigma$ are only included in the update for $w_i$, not in
the objective function for updating $\beta$ and $\Sigma$. In the Supplementary
Materials, we outline a procedure for obtaining starting values that can improve
computing times and the quality of the resulting estimates relative to naive
starting values.

\section{Approximate likelihood ratio testing} \label{sec:testing}
To make inferences about the parameters we use the approximate negative
log-likelihood $h_n(\beta, \Sigma \mid w_1, \dots, w_n)$, where $w_1, \dots,
w_n$ are held at the final iterates given by Algorithm 1. Focus is on
testing hypotheses of the form $(\beta, \Sigma) \in \mathbb{H}_0$
versus $(\beta, \Sigma) \in \mathbb{H}_A$, where $\mathbb{H}_0$ and
$\mathbb{H}_A$ partition the parameter set. If the null hypothesis constrains $\beta$ only, we write for simplicity $\beta
\in \mathbb{H}_0$, and similarly with $\Sigma \in \mathbb{H}_0$ if the null hypothesis constrains
$\Sigma$ only. We propose the test statistic $ T_n
= h_n(\tilde{\beta}, \tilde{\Sigma}\mid \tilde{w}_1, \dots, \tilde{w}_n) -
h_n(\bar{\beta}, \bar{\Sigma}\mid \tilde{w}_1, \dots, \tilde{w}_n)$, where
$(\tilde{\beta}, \tilde{\Sigma})$ and the approximation points $\tilde{w} =
\{\tilde{w}_1, \dots, \tilde{w}_n\}$ are obtained by running Algorithm 1 with
the restrictions implied by $\mathbb{H}_0$ and $(\bar{\beta}, \bar{\Sigma}) =
\argmin_{(\beta, \Sigma) \in \mathbb{H}_0\cup \mathbb{H}_A} h_n(\beta, \Sigma
\mid \tilde{w}_1, \dots, \tilde{w}_n)$. That is, $(\tilde{\beta},
\tilde{\Sigma})$ and $\tilde{w}$ are estimates and expansion points,
respectively, from fitting the null model while $\bar{\beta}$ and $\bar{\Sigma}$
are obtained by maximizing the working likelihood from \eqref{eq:work:model}
with the expansion points fixed at those obtained by fitting the null model. We
fix the expansion points to ensure $(\tilde{\beta}, \tilde{\Sigma})$ and
$(\bar{\beta}, \bar{\Sigma})$ are maximizers of the same working likelihood, but
under different restrictions. We chose the null model's expansion points to be
conservative; that is, to favor the null hypothesis model. If the working model
is accurate and $\mathbb{H}_0$ contains no boundary points of the parameter set,
we expect $T_n$ to be, under the null hypothesis, approximately chi-square
distributed with degrees of freedom equal to the number of
restrictions implied by $\mathbb{H}_0$.

A main motivation for our method is inference on the covariance matrix $\Sigma$.
Null models corresponding to hypotheses that constrain elements of $\Sigma$ are
straightforward to fit by including those constraints in the definition of the
set $\mathbb{M}$ in the update for $\Sigma$. For example, to test whether the
first and second element of $Y$ are independent, fitting the null model
corresponds to setting
\[
\mathbb{M} = \{\Sigma \in \SPSD{r} : \Sigma_{12} = \Sigma_{21} = 0\},
\]
assuming there are no other restrictions. Similarly, independence of more than
two responses can be tested by including more off-diagonal restrictions in the
definition of $\mathbb{M}$, and equality of variances for some responses can be
tested by including restrictions such as $\Sigma_{11} = \Sigma_{22}$. In
principle our method could also be used to test restrictions such as
$\Sigma_{11} = 0$, corresponding to the first latent variable being constant.
However, any null hypothesis which forces some eigenvalues of $\Sigma$ to be
zero corresponds to testing of boundary points, and then the likelihood ratio
test-statistic has a different asymptotic
distribution \citep{Self.Liang1987,Geyer1994,Baey.etal2019}.

A formal statement of the full
algorithm for hypothesis testing is given in Algorithm 2. We investigate the
size and power of the proposed procedure in Section \ref{sec:LRT}.

\begin{algorithm}[t]
\noindent \textbf{Algorithm 2}: Hypothesis testing procedure for $(\beta,
\Sigma) \in \mathbb{H}_0$ versus $(\beta, \Sigma)\in \mathbb{H}_A$
\begin{enumerate}
\item Given $\mathbb{H}_0$ and $\mathbb{H}_A,$ initialize
$(\beta^{(1)}, \Sigma^{(1)}) \in \mathbb{H}_0$.
\item For $k = 1, 2, \dots$ until convergence:
\begin{enumerate}
\item $w_i^{(k+1)} = \argmax_{w \in \mathbb{R}^r} \left\{\log
f_{\beta^{(k)}, \underline{\Sigma}^{(k)}}(y_i, w) - \tau \Vert y_i -
X_i\beta^{(k)}\Vert^2\right\}$ for $i = 1, \dots, n$
\item $(\beta^{(k+1)}, \Sigma^{(k+1)}) = \argmin_{(\beta, \Sigma) \in
\mathbb{H}_0} h_n(\beta, \Sigma\mid w_1^{(k+1)},
\dots, w_n^{(k+1)})$\hfill
\end{enumerate}
\item Set $(\tilde{\beta}, \tilde{\Sigma}) = (\beta^{(k^*)}, \Sigma^{(k^*)})$
and $\tilde{w}_i = w_i^{(k^*)}$ for $i = 1, \dots, n$ where $k^*$ denotes the
final iterate from 2.
\item Compute $(\bar{\beta}, \bar{\Sigma}) = \argmin_{(\beta, \Sigma) \in \mathbb{H}_0 \cup
\mathbb{H}_A} h_n(\beta, \Sigma\mid \tilde{w}_1,
\dots, \tilde{w}_n)$
\item Return $T_n = h_n(\tilde{\beta}, \tilde{\Sigma}\mid \tilde{w}_1,
\dots, \tilde{w}_n) - h_n(\bar{\beta}, \bar{\Sigma} \mid\tilde{w}_1, \dots,
\tilde{w}_n).$
\end{enumerate}
\end{algorithm}

\section{Numerical experiments} \label{sec:simulations}

\subsection{Overview}

We are not aware of a publicly available (or otherwise) software that fits our
model outside of special cases. We therefore compare to existing methods which assume related but
somewhat different models. Specifically, when investigating prediction (Section
\ref{sec:sim:pred}) and estimation of $\beta$ (Supplementary Materials) we
compare to separate GLMMs for the different response types. We chose this
comparison because the GLMMs have correctly specified marginal distributions in
our setting. In particular, the marginal moments $\E(Y_j)$ and $\var(Y_j)$, $j =
1, \dots, r$, are correctly specified under the GLMMs. Thus, we expect the GLMMs
to give reasonable estimates of the mean functions, and expect differences in
predictive performance to our method to be indicative of the usefulness of joint
modeling and estimating the off-diagonal elements of $\Sigma$. We also compare
our method on prediction, estimation of $\beta$, and estimation of $\cor(Y)$ to
multivariate covariance generalized linear models (MCGLMs)
\citep{Bonat.Jorgensen2016}. MCGLMs model the marginal moments of $Y$ and,
accordingly, provide estimates of the mean vector and covariance matrix of $Y$,
but not of $\Sigma$. To further highlight the usefulness
of modeling dependence, we include results for our method with $\Sigma$
constrained to be diagonal, i.e., with responses assumed independent. Whether
constraining $\Sigma$ to be diagonal or not, diagonal elements corresponding to
Bernoulli responses are constrained to equal one when using our method.

\subsection{Prediction comparisons} \label{sec:sim:pred}

We consider $r = 9$ response variables, three normally distributed, three
Poisson, and three Bernoulli. When fitting separate GLMMs for the
three-dimensional response vectors with elements of the same type, we consider
two covariance structures that common software can fit: (i) $\Sigma_{jj} =
\sigma_j^2$ and $\Sigma_{jk} = 0$ for $j \neq k$ or (ii) $\Sigma = \sigma^2
1_31_3^\tsp$, for some $\sigma^2 > 0$ and $1_3 = [1, 1, 1]^\tsp$. Option (i)
assumes all responses are independent and option (ii) corresponds to using a
shared random effect for observations of the same type in the same response
vector. We refer to these as independent and clustered GLMMs, respectively.
There are many software packages for fitting GLMMs. We pick the \texttt{glmm}
\citep{Knudson.etal2021} package to fit the independent GLMMs and fit the
clustered GLMMs using \texttt{lme4} \citep{Bates.etal2015}. Briefly, the
former uses a Monte Carlo approximation of the likelihood and the latter uses
adaptive Gaussian quadrature.

Predictions are formed by plugging estimates from the different methods into
the expressions for $\E(Y_i) = \E\{\E(Y_i\mid W_i)\}$ in Section
\ref{sec:model}, and for simplicity we define prediction errors as the
differences between those expectations and the observed responses, regardless of
type. When a closed form expression is unavailable, the expectation is
obtained by $r$ one-dimensional numerical integrations. We compare to (oracle)
predictions using the true $\beta$ and $\Sigma$.

We next describe how data are generated in the simulations. The responses have
different predictors and $\beta$ is partitioned accordingly: $\beta = [\beta_1^\tsp,
\dots, \beta_r^\tsp]^\tsp$, $\beta_j \in \R{p_j}$,  and $q = \sum_{j = 1}^r p_j$.
We write $X_{i, j} \in \R{p_j}$ for the $i$th observation of the
predictors for the $j$th response. In all simulations, each $X_{i, j}$
consists of a one in the first element (an intercept) and, in the remaining
$p_j - 1$ elements, independent realizations of a ${\rm U}[-1,1]$ random
variable, where $p_j = p_k$ for all $j$ and $k$. For $j = 1,\dots,  r$, the
true regression coefficient $\beta_{j}$ has first element equal to
$\beta_{0j}$ and all other elements chosen as independent realizations of a
${\rm U}[-.5,.5]$. We set $\beta_{0j} = 2$ if the response is normal or
(quasi-)Poisson, and equal to zero if the response is Bernoulli. Similarly, if
the response is normal, we set $\psi_j = .01$; otherwise, we set $\psi_j = 1$.

We consider three different structures for $\Sigma$: for some $\rho \in (0,
1)$ we set $\Sigma = 0.5 \tilde{\Sigma}$ where $\tilde{\Sigma}$ is (i)
autoregressive ($\tilde{\Sigma}_{jk} = \rho^{|j-k|}$); (ii) compound symmetric
($\tilde{\Sigma}_{jk} = \rho \I(j\neq k) + \I(j = k)$); or (iii)
block-diagonal, meaning $\tilde{\Sigma}_{jk} = \rho \I(j \neq k) + \I(j = k)$
if $(j,k) \in \left\{1, 4, 7\right\} \times \left\{1, 4, 7\right\}$, $(j,k)
\in \left\{2,5,8\right\} \times \left\{2,5, 8\right\}$, or $(j,k) \in
\left\{3,6,9\right\} \times  \left\{3,6,9\right\}$ and zero otherwise. The
first through third responses are normal, the fourth through sixth Bernoulli,
and seventh through ninth Poisson. Hence, each of the blocks given by
the structure in (iii) includes one of each response type. These structures
are used to generate the data but are not imposed when fitting models. For all
the structures, the GLMMs have correctly specified diagonal elements of
$\Sigma$, but incorrectly specified off-diagonal elements in general.

For each structure of $\Sigma$, we investigate the effects of the sample size
($n$), the number of predictors ($p_j$, $j = 1, \dots, r$), and the
correlation parameter ($\rho$). We present relative squared out of sample prediction
errors, defined as the ratio of a method's sum of squared prediction
error to the sum of squared prediction error of the oracle prediction.
Averages are based on 500 independent replications, and for each replication,
out of sample predictions are on an independent test set of $10^4$
observations.

\begin{figure}[t]
 \centering
 \includegraphics[width=.95\textwidth]{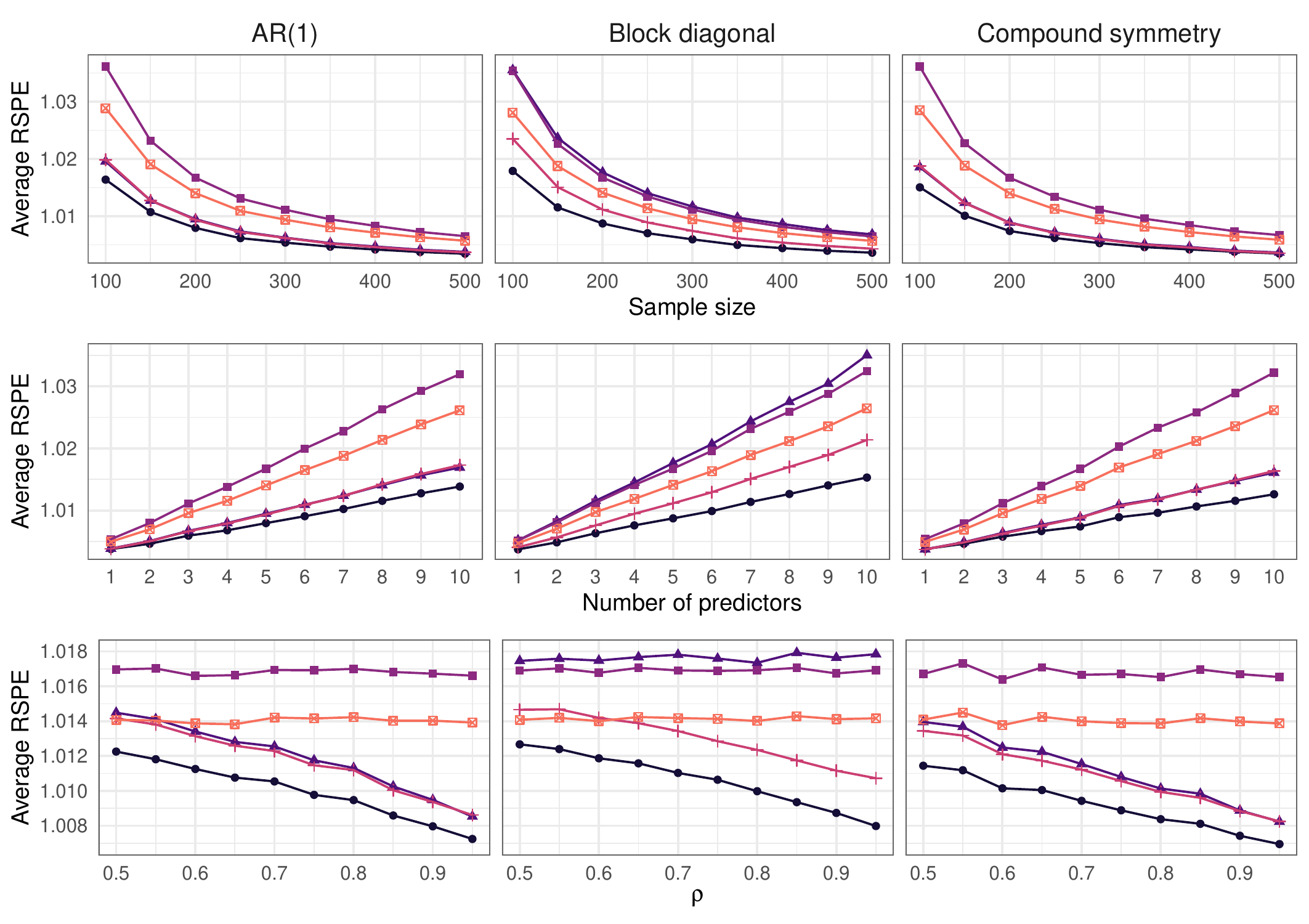}\\
 \includegraphics[width=.6\textwidth]{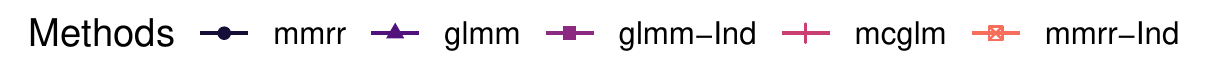}\\
 \caption{Average relative squared prediction errors. Top: $\rho = 0.9$ and $p_j = 5$ for $j = 1, \dots, 9$. Middle: $n= 200$ and $\rho = 0.9$. Bottom: $n = 200$ and $p_j = 5$ for $j = 1,
  \dots, 9$. glmm indicates clustered GLMMs, glmm-Ind GLMMs with diagonal covariance matrix, mcglm the method of \cite{Bonat.Jorgensen2016}, mmrr the proposed method, and mmrr-Ind the proposed method with diagonal covariance matrix.}
 \label{fig:AltMethods_Results}
 \end{figure}

In the top row of Figure \ref{fig:AltMethods_Results}, as $n$ increases, each
method's performance improves relative to oracle predictions. However, across
all settings, the proposed method performs best. When the covariance structure
is non-sparse (e.g., autoregressive or compound symmetric), the clustered
GLMMs can outperform our method with the diagonal covariance matrix and the
independent GLMMs. The same relative performances are observed as $p$
increases in the middle row; and when $\rho$ increases in the bottom row. When
$\Sigma$ is block-diagonal, both versions of our method outperform the
competitors. In the case of clustered GLMMs, this is likely due to the
specified covariance structure being a poor approximation to the true
covariance. For independent GLMMs, this is likely due to the fact that
\texttt{glmm} (nor other software) can impose the identifiability condition on
$\Sigma$ for the Bernoulli responses. MCGLMs perform second-best in most
settings, can perform similarly to culstered GLMMs when the true covariance
structure is non-sparse, and can be outperformed by our method with diagonal
covariance when dependence is weak. In summary, the results show  the
usefulness of joint modeling for prediction, and they illustrate the
usefulness of the proposed method relative to ones based on marginal moments.

The Supplementary Materials include results on mean squared errors for
estimating $\beta$, for the same methods and settings used in Figure
\ref{fig:AltMethods_Results}, and the results are qualitatively similar to
those in Figure \ref{fig:AltMethods_Results}. The Supplementary Materials also
include a comparison to separate GLMs, effectively assuming independence
between responses, which leads to substantially worse predictions than all
methods considered here.
\subsection{Performances for different response types}\label{sec:SUR_Sims}
In Figure \ref{fig:AltMethods_Results} averages were taken over all responses
types. To see if the benefits of joint modeling are greater for some response
types, it is of interest to stratify results by type. We compare the two
versions of our method (diagonal versus non-diagonal $\Sigma$). Because both
versions have correctly specified univariate response distributions and are fit
using the same algorithm except for the constraints on $\Sigma$, these
simulations investigate the usefulness of joint modeling of mixed-type
responses. Data are generated as in the previous section.

\begin{figure}[t]
\centering
\includegraphics[width=.95\textwidth]{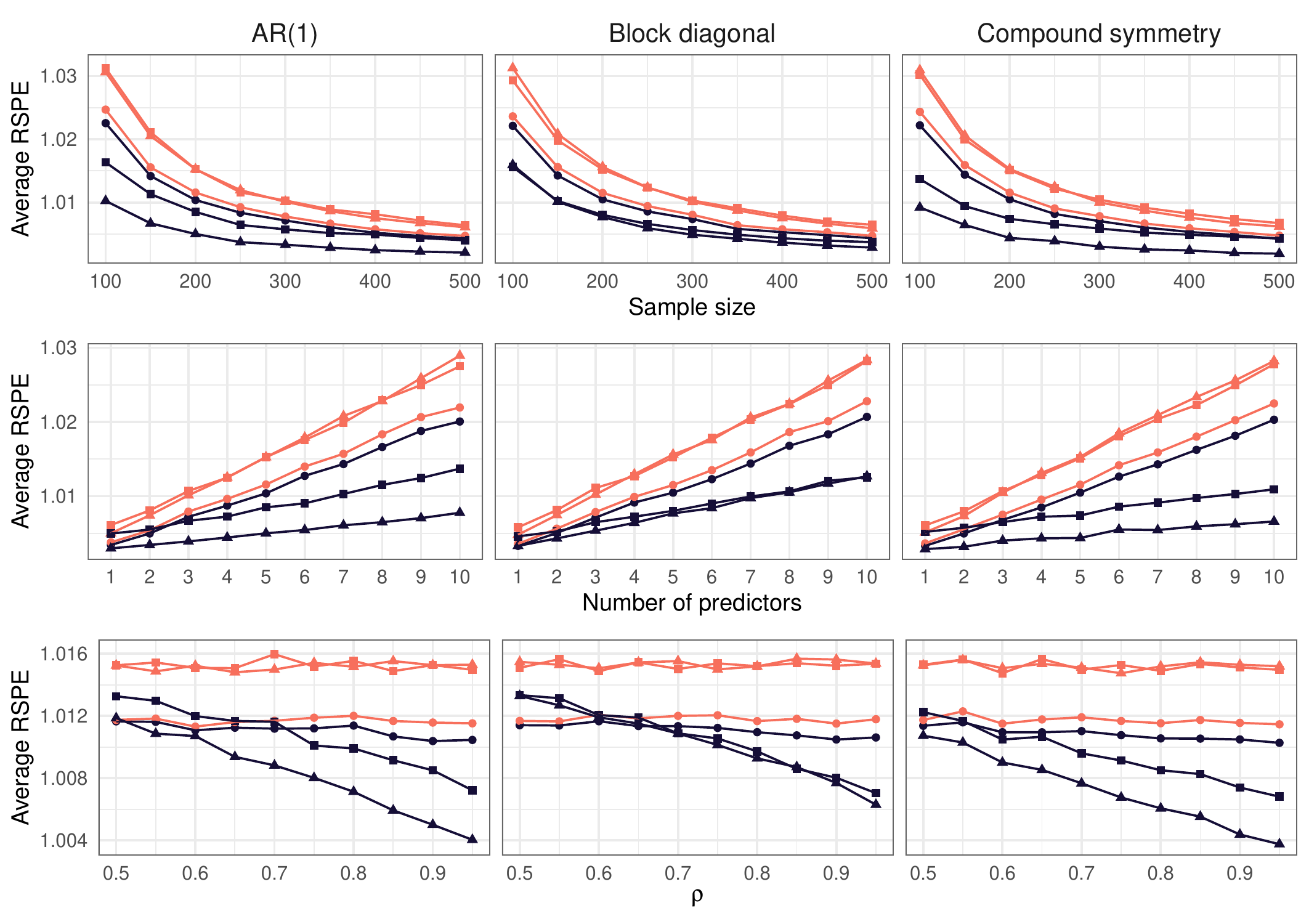}
  \includegraphics[width=.85\textwidth]{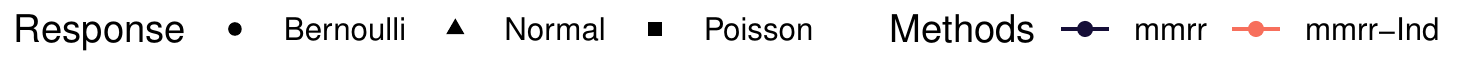}\\
\caption{Average relative squared prediction errors. Top: $\rho = 0.9$ and $p_j = 5$
for $j = 1, \dots, 9$. Middle: $n=
200$ and $\rho = 0.9$. Bottom $n = 200$
and $p_j = 5$ for $j = 1, \dots, 9$. mmrr is the proposed method and mmrr-Ind the proposed method with diagonal covariance matrix.} \label{fig:SUR_Sims}
\end{figure}

In the first row of Figure \ref{fig:SUR_Sims}, as $n$ increases, both methods'
relative mean squared prediction error approaches the oracle prediction error.
However, the predictions from joint modeling outperforms those using a diagonal
$\Sigma$. The differences between the two methods are smallest for Bernoulli
responses. A similar result is observed in the second row: as $p_j$ approaches
10, both methods' relative performance degrades, although for all three response
types, predictions from the joint modeling degrades more gradually. In
the bottom-most row, we display results as $\rho$ varies. When $\rho = 0.5$
there is a less substantial difference between the two methods. As $\rho$
approaches 0.95, the difference between the two methods becomes greater. This
result is also observed in the Bernoulli responses, but to a lesser degree than
the normal and Poisson.

In Section F.2 of the Supplementary Materials, we
present a simulation study which focuses on modeling many Bernoulli responses
and a single normal response. Those results highlight that even though the
relative squared prediction error for the Bernoulli responses is only slightly
improved by joint modeling, one can realize substantial prediction accuracy
gains for the single normal response variable by exploiting dependence between
it and the Bernoulli responses. We present similar results comparing our method
and MCGLMs in the Supplementary Materials.

\subsection{Covariance estimation} \label{sec:sim:cov}

To investigate the usefulness of the proposed method for estimating covariances
and correlations specifically, we consider a setting without predictors. That
is, we consider \eqref{eq:model} with $x = 1$, corresponding to an intercept
only. Ideally we would compare to an established method for estimating $\Sigma$
in our setting. Since none is available (to the best of our knowledge), we
compare the estimates of $\Omega = \cov(Y)$ and $\cor(Y) =
\diag(\Omega)^{-1/2}\Omega\diag(\Omega)^{-1/2}$ from our method to moment-based
estimates. One is the empirical (or sample) covariance matrix $S = n^{-1}\sum_{i
= 1}^n (y_i - \bar{y})(y_i - \bar{y})^\tsp$, where $\bar{y} = n^{-1}\sum_{i =
1}^n y_i$ is the sample mean. The corresponding empirical correlation matrix is
$\diag(S)^{-1/2} S \diag(S)^{-1/2}$. MCGLMs also provide a moment-based estimate
of $\cor(Y)$, which we found to be indistinguishable from the empirical
correlation matrix in the present setting without predictors.

We note the comparison to $S$, or other moment-based estimates of $\cov(Y)$ not
assuming \eqref{eq:model}, is not ideal because $S$ cannot in general be mapped
to an estimate of $\Sigma$. Formally the issue is that while $\Omega$ is
injective as a function of $\Sigma$, which follows from Theorem \ref{thm:id},
that function is not onto $\SPSD{r}$. Put differently, not all realizations of
$S$ give an estimate consistent with \eqref{eq:model}. Nevertheless, $S$ is
(strongly) consistent for $\Omega$ when there are no predictors since the
observations are then i.i.d., so we expect reasonable estimates of $\Omega$.

\begin{figure}[h]
\centering
  \includegraphics[width=\textwidth]{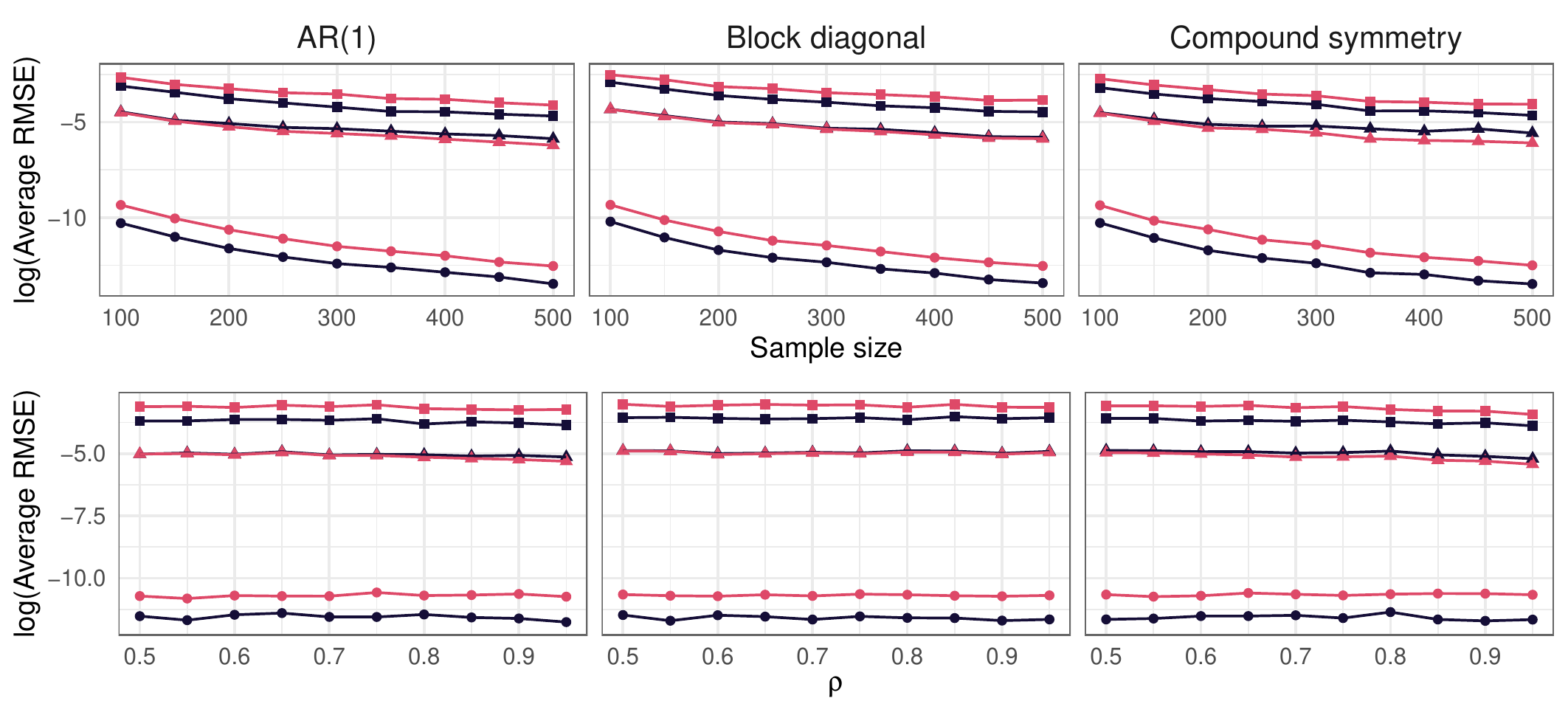}
   \includegraphics[width=.85\textwidth]{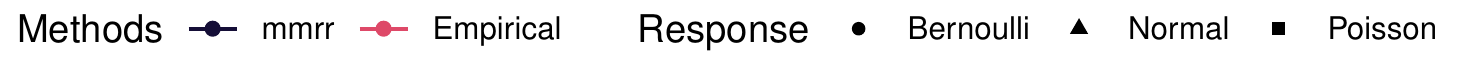}
  \caption{Average relative mean squared error for estimating the diagonals of ${\rm cov}(Y)$ stratified by response type. (Top row) Results with $\rho= 0.9.$ (Bottom row) Results with $n = 200$. mmrr is the proposed method and Empirical the sample covariance.}
  \label{fig:Marginal_Cov1}
\end{figure}

The data generating model is as in Section \ref{sec:sim:pred}, but with an
intercept only. Figure \ref{fig:Marginal_Cov1} shows average relative mean
square error for the diagonal entries of $\Omega$ by type. Specifically, we
report the averages of: (normal) $\sum_{j=1}^3 (\Omega_{jj} -
\hat{\Omega}_{jj})^2/\sum_{j=1}^3\Omega_{jj}^2$, (Bernoulli) $\sum_{j=4}^6
(\Omega_{jj} - \hat{\Omega}_{jj})^2/\sum_{j=4}^6\Omega_{jj}^2$, and (Poisson)
$\sum_{j=7}^9 (\Omega_{jj} - \hat{\Omega}_{jj})^2/\sum_{j=7}^9\Omega_{jj}^2$
where $\hat{\Omega}$ is an estimate of $\Omega$ with $j$th diagonal entry
$\Omega_{jj}.$ In Figure \ref{fig:Marginal_Cov1}, we see our method
estimates the variances of the Poisson components more accurately than does
the sample covariance, but the two perform similarly for normal and Bernoulli
responses.

\begin{figure}[h]
\centering
  \includegraphics[width=\textwidth]{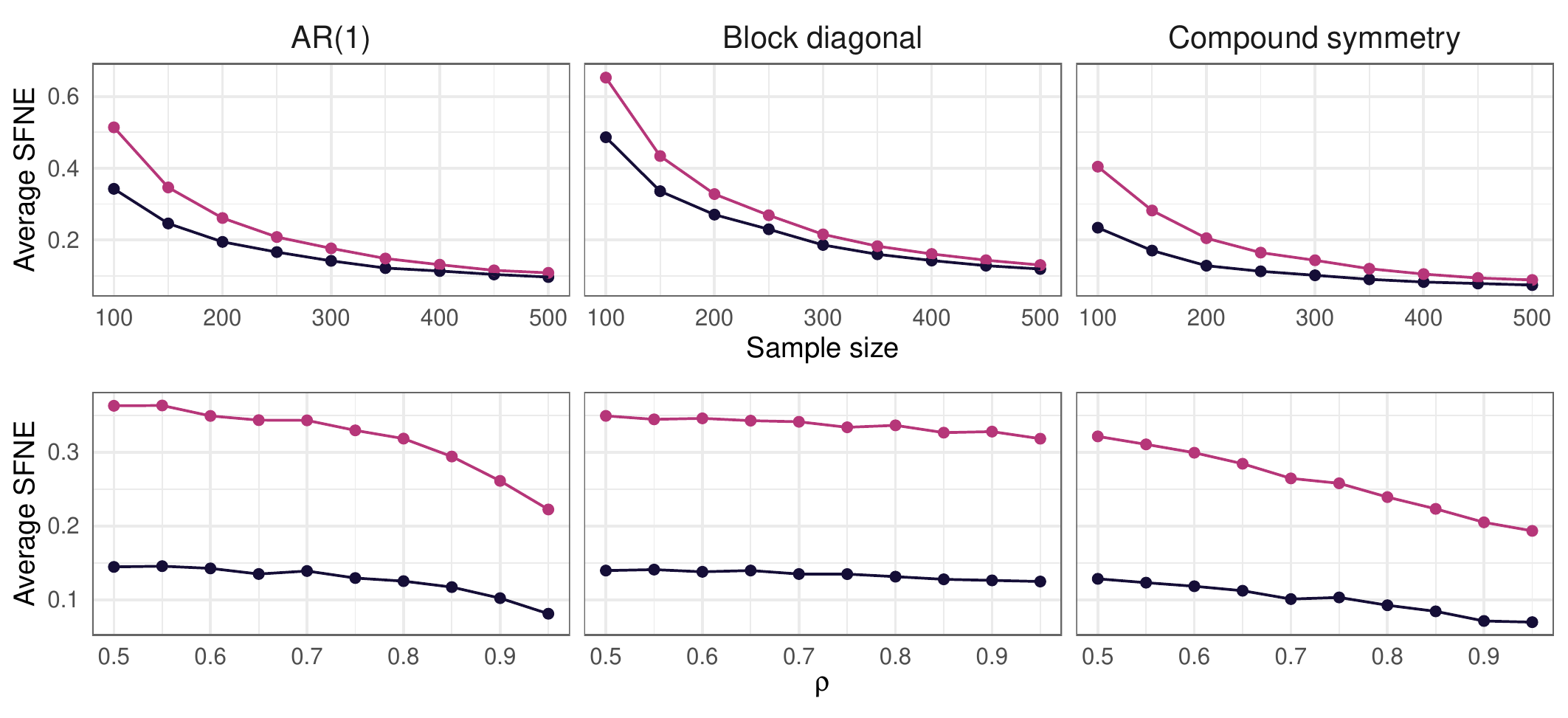}
   \includegraphics[width=.35\textwidth]{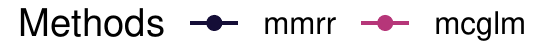}
  \caption{Average squared Frobenius norm error for estimating ${\rm cor}(Y)$. (Top row) Results with $\rho= 0.9.$ (Bottom row) Results with $p = 1$ and $n = 200$. mmrr is the proposed method and mcglm the method of \citet{Bonat.Jorgensen2016}.}
  \label{fig:Marginal_Cor1}
\end{figure}

Figure \ref{fig:Marginal_Cor1} displays the average mean squared estimation
error for $\diag(\Omega)^{-1/2}\Omega\diag(\Omega)^{-1/2}$, the correlation
matrix of $(Y_1, \dots, Y_9)^\top$. For smaller sample sizes,
the proposed method performs better than MCGLMs. For larger sample sizes the
differences diminish somewhat, which is not surprising given that sample
correlations are consistent. As the correlation parameter $\rho$ varies with $n
= 200$, we see that the differences between the methods remain relatively
constant, with the proposed one being better in every setting.

\subsection{Approximate likelihood ratio testing}\label{sec:LRT}
We examine the approximate likelihood ratio testing procedure described in
Section \ref{sec:testing}. Let $\DPD{r}$ be the set of $r\times r$ diagonal and
positive definite matrices. We study the size and power of the proposed tests
for $\Sigma \in \mathbb{H}_0 = \DPD{r}$ versus $\Sigma \in \mathbb{H}_A =
\SPD{r}\setminus\mathbb{H}_0$, and, assuming all responses have the same
predictors as in \eqref{eq:model}, $\Beta \in \mathbb{H}_0 = \{\Beta \in
\R{p\times r}: \Beta_{kj} = 0, ~j = 1, \dots, r\}$ versus $\Beta \in
\mathbb{H}_A = \R{p \times r}\setminus \mathbb{H}_0$, where $\Beta_{kj}$ denotes
the $k$th predictor's effect on the $j$th response variable, i.e., the $(k,j)$th
element of $\Beta$. We set $k=2$. Thus, the null hypothesis implies the first
predictor (ignoring the intercept) has no effect on any response. Multiple
testing corrections, which are often needed when using separate models for the
$r$ responses, are not needed here.

\begin{figure}[t]
\centering
  \centerline{\hfill\includegraphics[width=.95\textwidth]{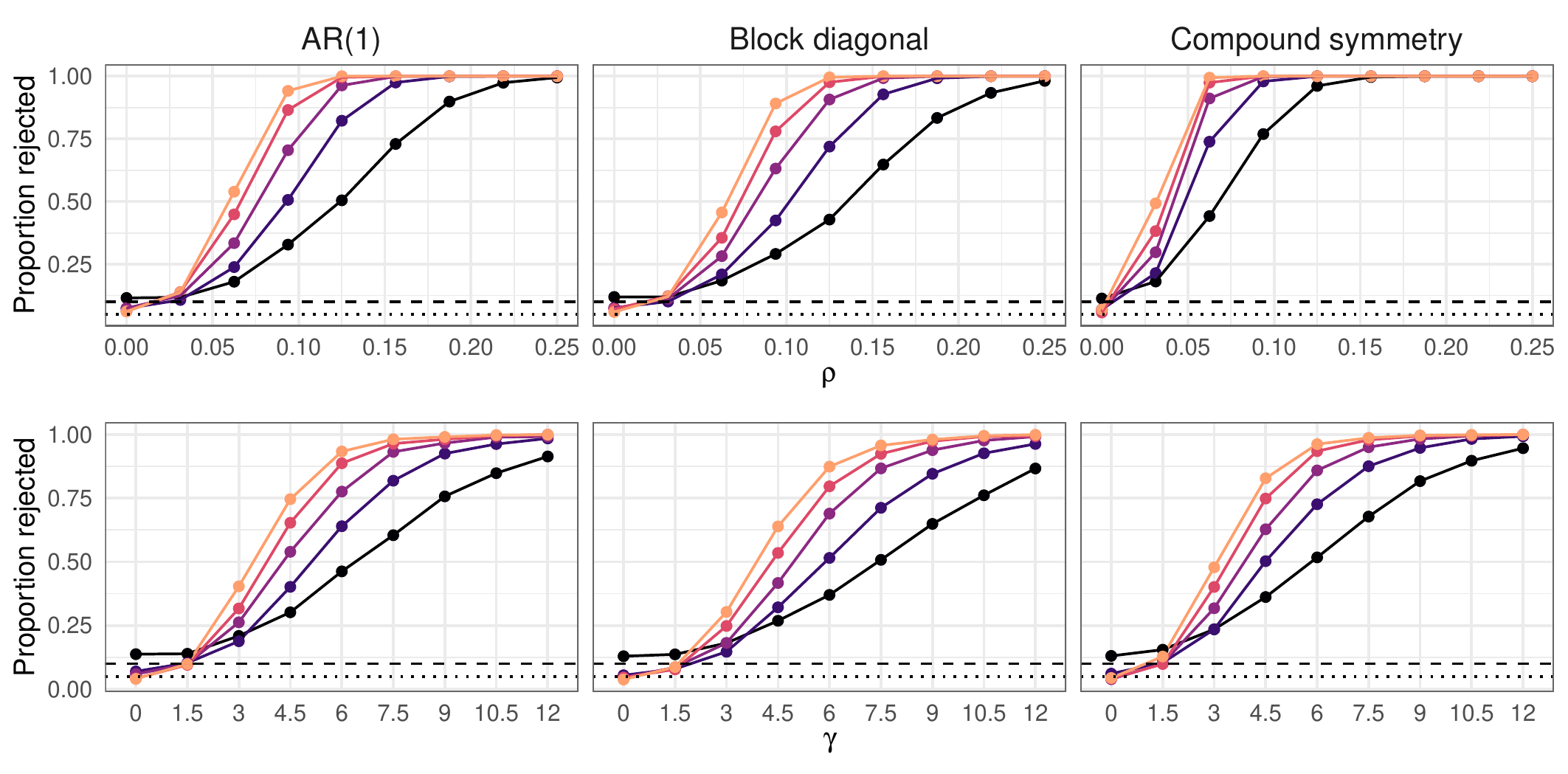}\hfill}
    \includegraphics[width=.5\textwidth]{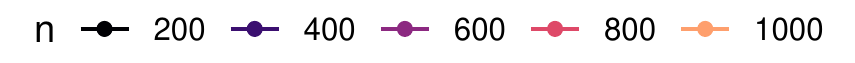}\\
  \caption{Top: Proportion of $\Sigma \in \mathbb{H}_0 = \DPD{r}$ rejected at 0.05 level. Bottom: Proportion of $\Beta \in \mathbb{H}_0 = \{\Beta \in
  \R{p\times r}: \Beta_{kj} = 0, ~j = 1, \dots, r\}$ rejected at the 0.05 level. Horizontal dashed and
  dotted black lines indicate 0.10 and 0.05.}
  \label{fig:LRT_Sims}
\end{figure}

Data are generated as in Section \ref{sec:SUR_Sims} but with $X_{i, 1} = X_{i,
2} = \dots = X_{i, r}$ for all $i=1, \dots, n$ and $\Beta = [\beta_1, \dots,
\beta_r] \in \R{p\times r}$. In the first setting, $n\in \{200, 400, \dots, 1000\}$ and
$\tilde\Sigma_{jk} = \rho^{|j-k|}$, $\rho \in \{0.0, 0.05, \dots, 0.4\}$. The
top row of Figure \ref{fig:LRT_Sims} displays the proportion of rejections at
the 0.05 significance level. When $\rho = 0$ (null hypothesis is true), the proportion of
rejections is approximately 0.10 when $n=200$, below $0.075$ when $n\geq 400$,
and near 0.05 (the nominal level) when $n = 2000$. As $\rho$ increases, even
with $n = 200$, the proportion of correctly rejected null hypotheses is near one
when $\rho = 0.4$. The power depends positively on both the magnitude of $\rho$
and the sample size.

In the second setting, we fix $\rho = 0.5$ and study how the effect size of the
$\Beta_{kj}$ affects power. After generating $\Beta$ as in Section
\ref{sec:SUR_Sims}, for $j = 1, \dots, r$ independently, we replace $\Beta_{kj}$
with a realization of a ${\rm U}[-\gamma 10^{-2}, \gamma 10^{-2}]$ where
$\gamma \in [0,12]$. The second row of Figure \ref{fig:LRT_Sims} shows that when
$\gamma=0$, so that $\Beta_{kj} = 0$ for all $j = 1,\dots, r$, the proportion of
rejections is slightly above 0.10 for $n = 200$, but close to 0.05 (the nominal
size) for all $n \geq 400$. There is also an indication that correlation between
responses benefits power. For example, the power curves under compound symmetry
tend to be above the corresponding ones under block-diagonal structure.

\section{Data examples}\label{sec:data_ex}

\subsection{Fertility data}\label{subsec:Fertility}
We consider a dataset collected on 333 women who were having difficulty becoming
pregnant \citep{Cannon.etal2013}. The goal is to model four mixed-type response
variables related to the ability to conceive. The predictors are age and
three variables related to antral follicles: small antral follicle count,
average antral follicle count, and maximum follicle stimulating hormone level.
Antral follicle counts can be measured via noninvasive ultrasound and are therefore
often used to model fertility. 

The response variables quantify the ability to conceive in different ways. Two
are approximately normally distributed (square-root estradiol level and
log-total gonadotropin level); and two are counts (number of egg cells and
number of embryos).  We modeled the latter using our model with conditional
quasi-Poisson distributions. We set $\psi_j = 10^{-2}$ for continuous responses
and $\psi_j = 10^{-1}$ for counts. To illustrate how the proposed methods can be
applied, we test the null hypothesis that $\Sigma$ is diagonal and find evidence against it (p-value $< 10^{-16}$) using the test described in Section
\ref{sec:LRT}. That is, there is evidence suggesting the four responses are not
independent given the predictors. Fitting the unrestricted model using our
software took less than three seconds on a laptop computer with 2.3 GHz 8-Core
Intel Core i9 processor. The hypothesis testing procedure took less than six
seconds on the same machine.

The estimated correlation matrices for the four observed responses,
$\widehat{\cor}(Y_i\mid X_i = \bar{X})$, and the latent variables,
$\widehat{\cor}(W_i\mid X_i)$, are, respectively,
\begin{equation*}
\left( \begin{array}{r r r r}
1.00 & 0.01 & -0.08 & -0.09\\
0.01 & 1.00 &  -0.03 &  -0.09 \\
-0.08 &  -0.03 &  1.00 &  0.69 \\
-0.09 & -0.09 &   0.69 &   1.00
\end{array}\right)
~~~~~~~~~\text{and}~~~~~~~~~
\left( \begin{array}{r r r r}
1.00 &  0.02 & -0.09 & -0.10\\
0.02 & 1.00 &  -0.04 &  -0.09 \\
-0.09 &  -0.04 &  1.00 &  0.74 \\
-0.10 & -0.09 &   0.74 &   1.00
\end{array}\right),
\end{equation*}
\vspace{1pt}
where the variable ordering is square-root estradiol level, log-total
gonadotropin level, number of egg cells, and number of embyros. The estimate of
$\cor(Y_i\mid X_i)$ is here evaluated at $\bar{X} = \sum_{i = 1}^n X_i/n$. The
estimates indicate substantial positive correlation between the number of egg
cells and number of embryos, whereas estradiol and gonadotropin levels appear
weakly negatively correlated with these two variables.

We also test whether the small antra follicle count is a significant predictor
of any of the responses after accounting for age, average antral
follicle count, and maximum follicle stimulating hormone level. The number of
small antra follicles (2-5 mm) is correlated with the number of total
antra follices (2-10 mm), and it has been argued that only total antra follicle
count are needed in practice \citep{LaMarca.Sunkara2013}. Fitting our model with
$\Sigma \in \mathbb{S}^{4}_{++}$, we reject the null hypothesis that the four
regression coefficients (one for each response) corresponding to antra follicle
count is zero (p-value $= 0.0052$).

Finally, to illustrate how uncertainty can be quantitified using the approximate
likelihood, we construct an approximate 95 \% confidence interval for
$\Sigma_{43}$ by inverting the proposed approximate likelihood ratio
test-statistic numerically. This gives the confidence interval $(0.17, 0.24)$,
which corresponds to correlations between $0.62$ and $0.8$. Confidence intervals
for the other parameters could be constructed similarly.

\subsection{Somatic mutations and gene expression in breast
cancer}\label{subsec:SomaticMutations}
We now focus on jointly modeling common somatic
mutations and gene expression measured on patients with breast cancer collected by
The Cancer Genome Atlas Project (TCGA). A somatic mutation is an alteration
in the DNA of a somatic cell. Somatic mutations are believed to play a central
role in the development of cancer. Because somatic mutations modify gene
expression, directly and indirectly, it is natural to model
somatic mutations and gene expression jointly.

The somatic mutation variables are binary, indicating presence or absence
of a somatic mutation in the region of a particular gene. We focus on the ten
genes where a somatic mutation was present in more than 5\% of subjects. Thus,
we have $r = 20$, coming from ten genes each with one response corresponding to
gene expression and one to the presence of a somatic mutation. For gene
expression, we model log-transformed RPKM measurements as normal random
variables. Each patient's age is included as a predictor.

\begin{figure}
\centering
\includegraphics[width=9cm]{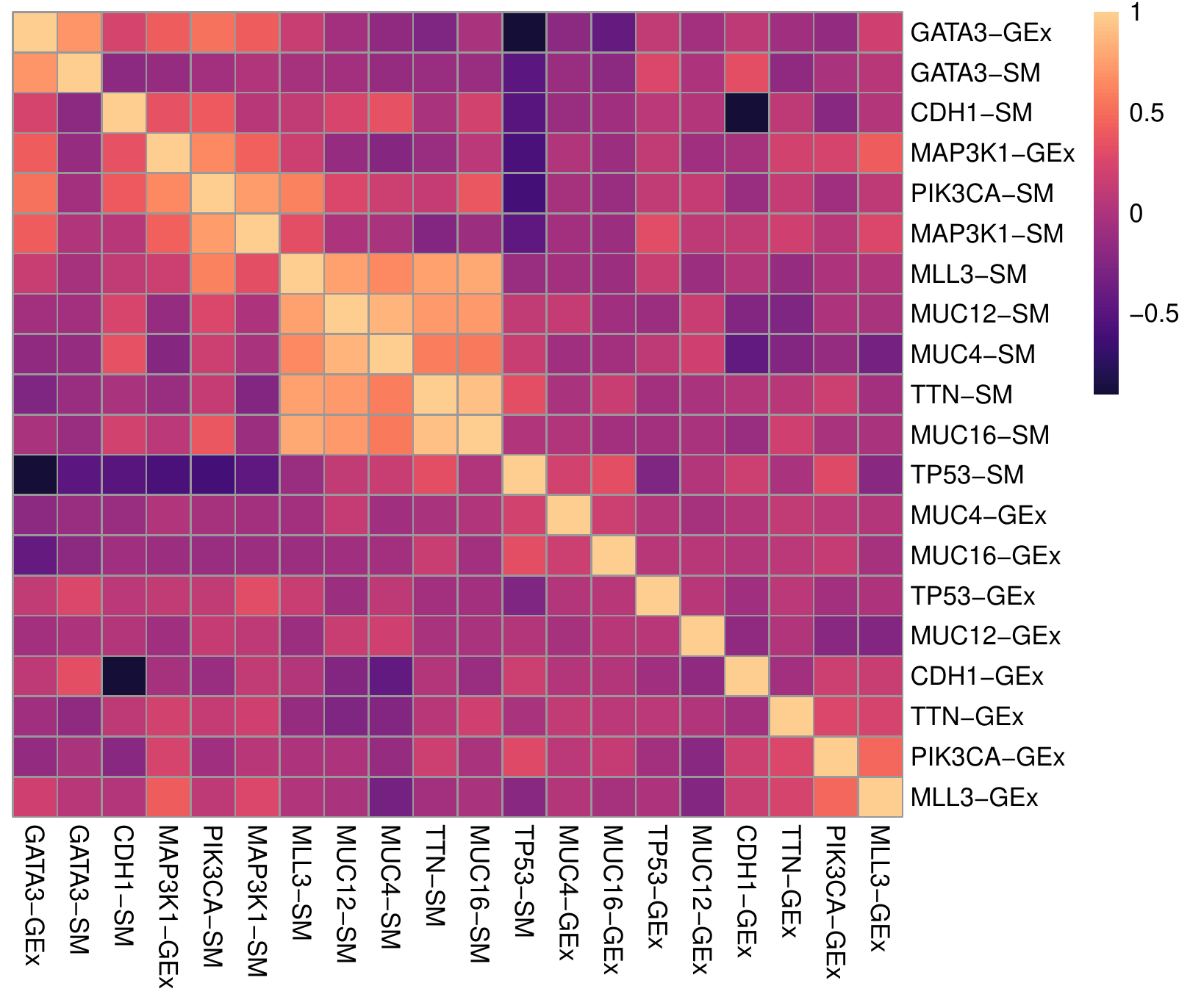}
\caption{Heatmap of the estimated correlation matrix for the $W_i \mid X_i$ in Section \ref{subsec:SomaticMutations}. Suffix -SM for somatic mutations; -GEx for gene expression.}\label{fig:SomaticMutations}
\end{figure}

We test the covariance matrix for block-diagonality. Under the null hypothesis, entries of
$\Sigma$ corresponding the correlations between somatic mutations and gene
expression measurements are zero (i.e., is no correlation between somatic
mutations and gene expression). The observed statistic is $T_n = 739$ with $100$
degrees of freedom for a p-value $< 10^{-16}.$ Figure \ref{fig:SomaticMutations}
displays the estimated correlation matrix for the $W_i \mid X_i$. We observe the
latent variables corresponding to somatic mutations and gene expression in CDH1
are highly negatively correlated, whereas for GATA3, somatic mutation and gene
expression latent variables have a strong positive correlation. Latent variables
for many of the somatic mutations are highly correlated (e.g., TTN, MLL3, MUC4,
MUC12, MUC16). However latent variables corresponding to some somatic mutations,
e.g., those in the region of TP53, exhibit small or even negative correlations
with many others (e.g., GATA3, CDH1, PIK3CA). Confidence intervals could be
constructed as in Example \ref{subsec:Fertility}.

\section{Discussion}

We have proposed a likelihood-based method for mixed-type multivariate response
regressions. Our
method gives an approximate maximum likelihood estimate of an unstructured
covariance matrix characterizing the dependence between responses. This is
particularly useful in settings where a dependence structure is not suggested by
subject-specific knowledge or one wants to discover such a structure using data.
To address the computational challenges with estimating an unstructured
covariance matrix, we have proposed a new algorithm. The algorithm handles
identifiability constraints
and it scales well in the dimension of the response vectors.

An advantage of likelihood-based methods compared to ones based on marginal
moments, such as for example generalized estimating equations and its
extensions \citep{Liang.Zeger1986,Rochon1996,Ziegler2011,Bonat.Jorgensen2016}, is
the plethora of existing procedures for inference and model selection using
likelihoods. For example, Wald tests and standard errors for the coefficient
estimates, based on the observed Fisher information, are readily available once
maximum likelihood estimates have been computed, as are information criteria. We
also used the likelihood to propose a testing procedure for the covariance
matrix. On the other hand, it is often computationally expensive to evaluate the
likelihood in latent variable models and it can lead to complicated
optimizations. We addressed that using a PQL-like approximation and a new
algorithm, and showed the resulting approximate maximum likelihood estimates are
often useful. Extending our method to other likelihood-approximations is a
possibility. For example, the PQL-approximation could be replaced by a Laplace
or Monte Carlo approximation. Another potential line of future research is the
asymptotic properties of PQL-type estimators, which despite the estimators'
apparent practical usefulness, are not fully understood.

Finally, we note some types of mixed-type longitudinal data are handled by our
method as-is, while other types would require modifications or would be better
analyzed using methods developed for that purpose. For example, our method can
be applied when the elements of $Y_i \in \R{r}$, $i = 1, \dots, n$, are repeated
measures of mixed-type responses on independently sampled patients indexed by
$i$. Modifications may be necessary if one wants to impose a particular
structure on $\Sigma$, if $n < r$, or if there is dependence between patients.

\subsection*{Supplementary Materials}
The Supplementary Materials include proofs, additional details, and an osteoarthritis initiative data analysis.
\vspace{-10pt}
\subsection*{Acknowledgements}
We thank three reviewers for suggestions which helped improve the manuscript. We
are grateful to Galin Jones and Adam Rothman for their comments on an earlier
version and  thank Charles McCulloch for sharing the OAI dataset analyzed in the
Supplementary Materials. Ekvall gratefully acknowledges support from FWF
(Austrian Science Fund) [P30690-N35]. Molstad's research was supported in part
by a grant from the National Science Foundation [DMS-2113589].

\bibliographystyle{apalike}
\bibliography{mixed_type}

\begin{thebibliography}{}

\bibitem[Baey et~al., 2019]{Baey.etal2019}
Baey, C., Courn{\`e}de, P.-H., and Kuhn, E. (2019).
\newblock Asymptotic distribution of likelihood ratio test statistics for
  variance components in nonlinear mixed effects models.
\newblock {\em Computational Statistics \& Data Analysis}, 135:107--122.

\bibitem[Bai et~al., 2020]{Bai.etal2020}
Bai, H., Zhong, Y., Gao, X., and Xu, W. (2020).
\newblock Multivariate mixed response model with pairwise composite-likelihood
  method.
\newblock {\em Stats}, 3(3):203--220.

\bibitem[Bates et~al., 2015]{Bates.etal2015}
Bates, D., M{\"a}chler, M., Bolker, B., and Walker, S. (2015).
\newblock Fitting linear mixed-effects models using lme4.
\newblock {\em Journal of Statistical Software}, 67(1).

\bibitem[Beck and Teboulle, 2009]{Beck.Teboulle2009}
Beck, A. and Teboulle, M. (2009).
\newblock A fast iterative shrinkage-thresholding algorithm for linear inverse
  problems.
\newblock {\em SIAM Journal on Imaging Sciences}, 2(1):183--202.

\bibitem[Bonat and J{\o}rgensen, 2016]{Bonat.Jorgensen2016}
Bonat, W.~H. and J{\o}rgensen, B. (2016).
\newblock Multivariate covariance generalized linear models.
\newblock {\em Journal of the Royal Statistical Society: Series C (Applied
  Statistics)}, 65(5):649--675.

\bibitem[Boyle and Dykstra, 1986]{Boyle.Dykstra1986}
Boyle, J.~P. and Dykstra, R.~L. (1986).
\newblock A method for finding projections onto the intersection of convex sets
  in {{Hilbert}} spaces.
\newblock In Brillinger, D., Fienberg, S., Gani, J., Hartigan, J., Krickeberg,
  K., Dykstra, R., Robertson, T., and Wright, F.~T., editors, {\em Advances in
  {{Order Restricted Statistical Inference}}}, volume~37, pages 28--47.
  {Springer New York}, {New York, NY}.

\bibitem[Breslow, 2004]{Breslow2004}
Breslow, N. (2004).
\newblock Whither {{PQL}}?
\newblock In {\em Proceedings of the second {{Seattle}} symposium in
  biostatistics}, pages 1--22. {Springer New York}.

\bibitem[Breslow and Clayton, 1993]{Breslow.Clayton1993}
Breslow, N.~E. and Clayton, D.~G. (1993).
\newblock Approximate inference in generalized linear mixed models.
\newblock {\em Journal of the American Statistical Association}, 88(421):9--25.

\bibitem[Cannon et~al., 2013]{Cannon.etal2013}
Cannon, A.~R., Cobb, G.~W., Hartlaub, B.~A., Legler, J.~M., Lock, R.~H., Moore,
  T.~L., Rossman, A.~J., and Witmer, J. (2013).
\newblock {\em Stat2: {{Building Models}} for a {{World}} of {{Data}}}.
\newblock {W. H. Freeman and Company}, {New York}.

\bibitem[Catalano, 1997]{Catalano1997}
Catalano, P.~J. (1997).
\newblock Bivariate modelling of clustered continuous and ordered categorical
  outcomes.
\newblock {\em Statistics in Medicine}, 16(8):883--900.

\bibitem[Catalano and Ryan, 1992]{Catalano.Ryan1992}
Catalano, P.~J. and Ryan, L.~M. (1992).
\newblock Bivariate latent variable models for clustered discrete and
  continuous outcomes.
\newblock {\em Journal of the American Statistical Association},
  87(419):651--658.

\bibitem[Cox and Wermuth, 1992]{Cox.Wermuth1992}
Cox, D.~R. and Wermuth, N. (1992).
\newblock Response models for mixed binary and quantitative variables.
\newblock {\em Biometrika}, 79(3):441--461.

\bibitem[{de Leon} and Carri{\`e}gre, 2007]{deLeon.Carriegre2007}
{de Leon}, A.~R. and Carri{\`e}gre, K.~C. (2007).
\newblock General mixed-data model: {{Extension}} of general location and
  grouped continuous models.
\newblock {\em Canadian Journal of Statistics}, 35(4):533--548.

\bibitem[Dunson, 2000]{Dunson2000}
Dunson, D.~B. (2000).
\newblock Bayesian latent variable models for clustered mixed outcomes.
\newblock {\em Journal of the Royal Statistical Society. Series B (Statistical
  Methodology)}, 62(2):355--366.

\bibitem[Ekvall and Jones, 2020]{Ekvall.Jones2020}
Ekvall, K.~O. and Jones, G.~L. (2020).
\newblock Consistent maximum likelihood estimation using subsets with
  applications to multivariate mixed models.
\newblock {\em Annals of Statistics}, 48(2):932--952.

\bibitem[Faes et~al., 2008]{Faes.etal2008}
Faes, C., Aerts, M., Molenberghs, G., Geys, H., Teuns, G., and Bijnens, L.
  (2008).
\newblock A high-dimensional joint model for longitudinal outcomes of different
  nature.
\newblock {\em Statistics in Medicine}, 27(22):4408--4427.

\bibitem[Fitzmaurice and Laird, 1995]{Fitzmaurice.Laird1995}
Fitzmaurice, G.~M. and Laird, N.~M. (1995).
\newblock Regression models for a bivariate discrete and continuous outcome
  with clustering.
\newblock {\em Journal of the American Statistical Association},
  90(431):845--852.

\bibitem[Fitzmaurice and Laird, 1997]{Fitzmaurice.Laird1997}
Fitzmaurice, G.~M. and Laird, N.~M. (1997).
\newblock Regression models for mixed discrete and continuous responses with
  potentially missing values.
\newblock {\em Biometrics}, 53(1):110--122.

\bibitem[Geyer, 1994]{Geyer1994}
Geyer, C.~J. (1994).
\newblock On the asymptotics of constrained {{M-estimation}}.
\newblock {\em Annals of Statistics}, 22(4):1993--2010.

\bibitem[Gueorguieva, 2001]{Gueorguieva2001}
Gueorguieva, R.~V. (2001).
\newblock A multivariate generalized linear mixed model for joint modelling of
  clustered outcomes in the exponential family.
\newblock {\em Statistical Modeling}, 1(3):177--193.

\bibitem[Gueorguieva and Agresti, 2001]{Gueorguieva.Agresti2001}
Gueorguieva, R.~V. and Agresti, A. (2001).
\newblock A correlated probit model for joint modeling of clustered binary and
  continuous responses.
\newblock {\em Journal of the American Statistical Association},
  96(455):1102--1112.

\bibitem[Gueorguieva and Sanacora, 2006]{Gueorguieva.Sanacora2006}
Gueorguieva, R.~V. and Sanacora, G. (2006).
\newblock Joint analysis of repeatedly observed continuous and ordinal measures
  of disease severity.
\newblock {\em Statistics in Medicine}, 25(8):1307--1322.

\bibitem[Jiang, 2007]{Jiang2007}
Jiang, J. (2007).
\newblock {\em Linear and {{Generalized Linear Mixed Models}} and {{Their
  Applications}}}.
\newblock Springer {{Series}} in {{Statistics}}. {Springer-Verlag}, {New York}.

\bibitem[Kang et~al., 2021]{Kang.etal2021}
Kang, X., Chen, X., Jin, R., Wu, H., and Deng, X. (2021).
\newblock Multivariate regression of mixed responses for evaluation of
  visualization designs.
\newblock {\em IISE Transactions}, 53(3):313--325.

\bibitem[Knudson et~al., 2021]{Knudson.etal2021}
Knudson, C., Benson, S., Geyer, C., and Jones, G. (2021).
\newblock Likelihood-based inference for generalized linear mixed models:
  {{Inference}} with the {{R}} package glmm.
\newblock {\em Stat}, 10(1):e339.

\bibitem[La~Marca and Sunkara, 2013]{LaMarca.Sunkara2013}
La~Marca, A. and Sunkara, S.~K. (2013).
\newblock Individualization of controlled ovarian stimulation in {{IVF}} using
  ovarian reserve markers: from theory to practice.
\newblock {\em Human reproduction update}, 20(1):124--140.

\bibitem[Liang and Zeger, 1986]{Liang.Zeger1986}
Liang, K.-Y. and Zeger, S.~L. (1986).
\newblock Longitudinal data analysis using generalized linear models.
\newblock {\em Biometrika}, 73(1):13--22.

\bibitem[McCullagh and Nelder, 1989]{McCullagh.Nelder1989}
McCullagh, P. and Nelder, J.~A. (1989).
\newblock {\em Generalized {{Linear Models}}}.
\newblock {Chapman \& Hall/CRC}, {Boca Raton, FL}.

\bibitem[Megginson, 1998]{Megginson1998}
Megginson, R.~E. (1998).
\newblock {\em An {{Introduction}} to {{Banach Space Theory}}}.
\newblock {Springer}, {New York, NY}.

\bibitem[Nocedal and Wright, 2006]{Nocedal.Wright2006}
Nocedal, J. and Wright, S. (2006).
\newblock {\em Numerical {{Optimization}}}.
\newblock {Springer-Verlag GmbH}, {New York, NY}.

\bibitem[Ochs et~al., 2014]{Ochs.etal2014}
Ochs, P., Chen, Y., Brox, T., and Pock, T. (2014).
\newblock {{iPiano}}: inertial proximal algorithm for nonconvex optimization.
\newblock {\em SIAM Journal on Imaging Sciences}, 7(2):1388--1419.

\bibitem[Olkin and Tate, 1961]{Olkin.Tate1961}
Olkin, I. and Tate, R.~F. (1961).
\newblock Multivariate correlation models with mixed discrete and continuous
  variables.
\newblock {\em Annals of Mathematical Statistics}, 32(2):448--465.

\bibitem[Pinheiro and Bates, 1996]{Pinheiro.Bates1996}
Pinheiro, J.~C. and Bates, D.~M. (1996).
\newblock Unconstrained parametrizations for variance-covariance matrices.
\newblock {\em Statistics and computing}, 6(3):289--296.

\bibitem[Poon and Lee, 1987]{Poon.Lee1987}
Poon, W.-Y. and Lee, S.-Y. (1987).
\newblock Maximum likelihood estimation of multivariate polyserial and
  polychoric correlation coefficients.
\newblock {\em Psychometrika}, 52(3):409--430.

\bibitem[{Rabe-Hesketh} et~al., 2002]{Rabe-Hesketh.etal2002}
{Rabe-Hesketh}, S., Skrondal, A., and Pickles, A. (2002).
\newblock Reliable estimation of generalized linear mixed models using adaptive
  quadrature.
\newblock {\em The Stata Journal}, 2(1):1--21.

\bibitem[{Rabe-Hesketh} et~al., 2004a]{Rabe-Hesketh.etal2004}
{Rabe-Hesketh}, S., Skrondal, A., and Pickles, A. (2004a).
\newblock Generalized multilevel structural equation modeling.
\newblock {\em Psychometrika}, 69(2):167--190.

\bibitem[{Rabe-Hesketh} et~al., 2004b]{Rabe-Hesketh.etal2004a}
{Rabe-Hesketh}, S., Skrondal, A., and Pickles, A. (2004b).
\newblock {{GLLAMM Manual}}.
\newblock {\em U.C. Berkeley Division of Biostatistics Working Paper Series}.

\bibitem[Rochon, 1996]{Rochon1996}
Rochon, J. (1996).
\newblock Analyzing bivariate repeated measures for discrete and continuous
  outcome variables.
\newblock {\em Biometrics}, 52(2):740--750.

\bibitem[Sammel et~al., 1997]{Sammel.etal1997}
Sammel, M.~D., Ryan, L.~M., and Legler, J.~M. (1997).
\newblock Latent variable models for mixed discrete and continuous outcomes.
\newblock {\em Journal of the Royal Statistical Society. Series B
  (Methodological)}, 59(3):667--678.

\bibitem[Schall, 1991]{Schall1991}
Schall, R. (1991).
\newblock Estimation in generalized linear models with random effects.
\newblock {\em Biometrika}, 78(4):719--727.

\bibitem[Self and Liang, 1987]{Self.Liang1987}
Self, S.~G. and Liang, K.-Y. (1987).
\newblock Asymptotic properties of maximum likelihood estimators and likelihood
  ratio tests under nonstandard conditions.
\newblock {\em Journal of the American Statistical Association},
  82(398):605--610.

\bibitem[Yang et~al., 2007]{Yang.etal2007}
Yang, Y., Kang, J., Mao, K., and Zhang, J. (2007).
\newblock Regression models for mixed {{Poisson}} and continuous longitudinal
  data.
\newblock {\em Statistics in Medicine}, 26(20):3782--3800.

\bibitem[Zellner, 1962]{Zellner1962}
Zellner, A. (1962).
\newblock An efficient method of estimating seemingly unrelated regressions and
  tests for aggregation bias.
\newblock {\em Journal of the American Statistical Association},
  57(298):348--368.

\bibitem[Ziegler, 2011]{Ziegler2011}
Ziegler, A. (2011).
\newblock {\em Generalized {{Estimating Equations}}}.
\newblock Number 204 in Lecture notes in statistics. {Springer}, {New York}.

\end{thebibliography}

\end{document}